\documentclass{emulateapj}
\usepackage{natbib}
\usepackage{graphicx}
\usepackage{dcolumn}
\usepackage{bm}
\usepackage{subfigure}
\usepackage{wasysym}
\usepackage{verbatim}
\usepackage{xcolor}
\usepackage{color, colortbl}
\usepackage{aas_macros}
\usepackage{placeins}
\usepackage{wrapfig}
\usepackage{wrapfig}
\usepackage{hyperref}
\hypersetup{
    colorlinks=true,
    linkcolor=blue,
    filecolor=magenta,     
    citecolor=blue,
    urlcolor=cyan,
    pdftitle={Overleaf Example},
    pdfpagemode=FullScreen,
    }

\definecolor{Gray}{gray}{0.9}

\newcommand{\planck}{\textit{Planck}}

\begin{document}

\title{The Atacama Cosmology Telescope: limits on dark matter--baryon interactions from DR4 power spectra}
\author{Zack Li\altaffilmark{1},
Rui An\altaffilmark{2},
Vera Gluscevic\altaffilmark{2},
Kimberly K.~Boddy\altaffilmark{3},
J~Richard~Bond\altaffilmark{1},
Erminia Calabrese\altaffilmark{4},
Jo Dunkley\altaffilmark{5,6},
Patricio~A.~Gallardo\altaffilmark{7},
Yilun Guan\altaffilmark{8},
Adam Hincks\altaffilmark{9,10},
Kevin~M.~Huffenberger\altaffilmark{11},
Arthur Kosowsky\altaffilmark{12},
Thibaut Louis\altaffilmark{13},
Mathew S.~Madhavacheril\altaffilmark{14},
Kavilan~Moodley\altaffilmark{15,16},
Lyman~A.~Page\altaffilmark{5},
Bruce Partridge\altaffilmark{17},
Frank~J.~Qu\altaffilmark{18},
Maria Salatino\altaffilmark{19,20},
Blake Sherwin\altaffilmark{18},
Crist\'obal Sif\'on\altaffilmark{21},
Cristian Vargas\altaffilmark{22},
Edward J.~Wollack\altaffilmark{23}
}

\altaffiltext{1}{Canadian Institute for Theoretical Astrophysics, University of Toronto, Toronto, ON, Canada M5S 3H8}
\altaffiltext{2}{Department of Physics \& Astronomy, University of Southern California, Los Angeles, CA, 90007, USA}
\altaffiltext{3}{Department of Physics, The University of Texas at Austin, Austin, TX 78712, USA}
\altaffiltext{4}{School of Physics and Astronomy, Cardiff University, The Parade, Cardiff, CF24 3AA, UK}
\altaffiltext{5}{Joseph Henry Laboratories of Physics, Jadwin Hall, Princeton University, Princeton, NJ 08544}
\altaffiltext{6}{Department of Astrophysical Sciences, Princeton University, Princeton, New Jersey 08544, USA}
\altaffiltext{7}{Kavli Institute for Cosmological Physics, University of Chicago, Chicago, IL 60637, USA}
\altaffiltext{8}{Dunlap Institute for Astronomy and Astrophysics, University of Toronto, 50 St. George St., Toronto, ON M5S 3H4, Canada}
\altaffiltext{9}{David A. Dunlap Department of Astronomy \& Astrophysics, University of Toronto, 50 St. George St., Toronto, ON M5S 3H4, Canada}
\altaffiltext{10}{Specola Vaticana (Vatican Observatory), V-00120 Vatican City State}
\altaffiltext{11}{Department of Physics, Florida State University, Tallahassee FL, USA 32306}
\altaffiltext{12}{Department of Physics and Astronomy, University of Pittsburgh, Pittsburgh, PA 15260, USA}
\altaffiltext{13}{Universit\'e Paris-Saclay, CNRS, Institut d'astrophysique spatiale, 91405, Orsay, France.}
\altaffiltext{14}{Perimeter Institute for Theoretical Physics, 31 Caroline Street N, Waterloo ON N2L 2Y5, Canada}
\altaffiltext{15}{Astrophysics Research Centre, University of KwaZulu-Natal, Westville Campus, Durban 4041, South Africa}
\altaffiltext{16}{School of Mathematics, Statistics \& Computer Science, University of KwaZulu-Natal, Westville Campus, Durban 4041, South Africa}
\altaffiltext{17}{Department of Physics and Astronomy, Haverford College, Haverford, PA, USA 19041}
\altaffiltext{18}{DAMTP, Centre for Mathematical Sciences, University of Cambridge, Wilberforce Road, Cambridge CB3 OWA, UK}
\altaffiltext{19}{Department of Physics, Stanford University, Stanford, California 94305, USA}
\altaffiltext{20}{Kavli Institute for Particle Astrophysics and Cosmology, Stanford, CA 94305, USA}
\altaffiltext{21}{Instituto de F\'isica, Pontificia Universidad Cat\'olica de Valpara\'iso, Casilla 4059, Valpara\'iso, Chile}
\altaffiltext{22}{Instituto de Astrof\'isica and Centro de Astro-Ingenier\'ia, Facultad de F\'isica, Pontificia Universidad Cat\'olica de Chile, Av. Vicu\~na Mackenna 4860, 7820436 Macul, Santiago, Chile}
\altaffiltext{23}{NASA Goddard Space Flight Center, 8800 Greenbelt Rd, Greenbelt, MD 20771, USA}


\begin{abstract}
Diverse astrophysical observations suggest the existence of cold dark matter that interacts only gravitationally with radiation and ordinary baryonic matter.
Any nonzero coupling between dark matter and baryons would provide a significant step towards understanding the particle nature of dark matter. 
Measurements of the cosmic microwave background (CMB) provide constraints on such a coupling that complement laboratory searches.
In this work we place upper limits on a variety of models for dark matter elastic scattering with protons and electrons by combining large-scale CMB data from the \planck{} satellite with small-scale information from Atacama Cosmology Telescope (ACT) DR4 data. In the case of velocity-independent scattering, we obtain bounds on the interaction cross section for protons that are 
40\% tighter than previous constraints from the CMB anisotropy. For some models with velocity-dependent scattering we find best-fitting cross sections with a 2$\sigma$ deviation from zero, but these scattering models are not statistically preferred over $\Lambda$CDM in terms of model selection.
\end{abstract}
\maketitle

\section{Introduction}
\setcounter{footnote}{0}

Measurements of the cosmic microwave background (CMB) anisotropies let us view the early Universe as a high-energy gravitational laboratory. In the standard model of cosmology, the dynamics of the early Universe were dominated by scattering between radiation and baryonic matter, as well as gravitational interactions from dark matter, leading to the oscillations that generated the famous acoustic features in the CMB power spectrum. Precise measurements of these features have helped build a successful, predictive model for the contents, geometry, and evolution of the early Universe. The lack of deviations from this standard model of cosmology has provided stringent constraints on extensions that would change the CMB acoustic features, such as physics arising from neutrinos and axions \citep{boyarsky2009, brust2013, marsh2016, green2019, brinckmann2019}, interactions in the dark sector \citep{cyrracine2014, hlozek2017, buenabad2018}, and models of dark energy.

\begin{figure}[t]
\includegraphics[width=\columnwidth]{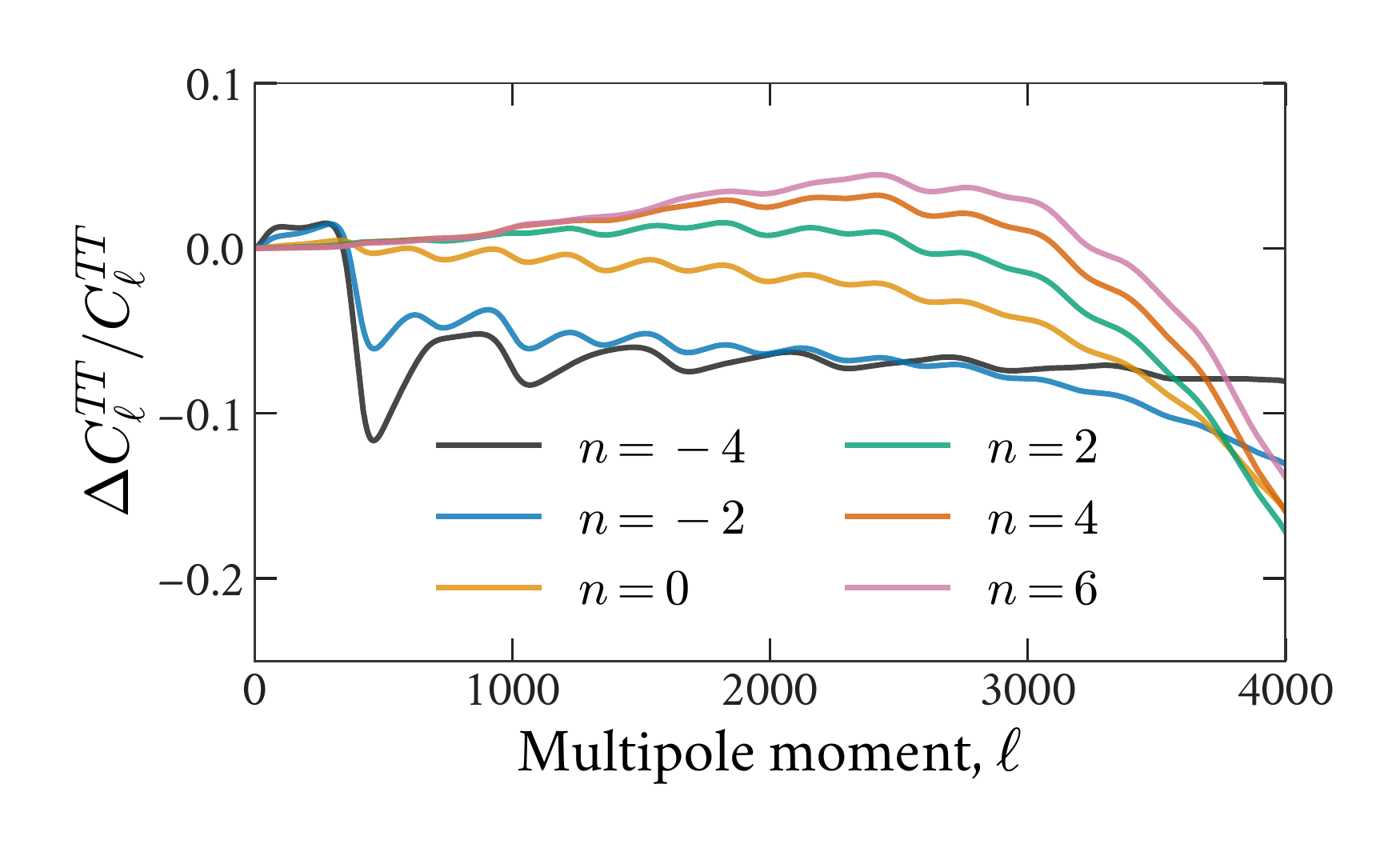}
\includegraphics[width=\columnwidth]{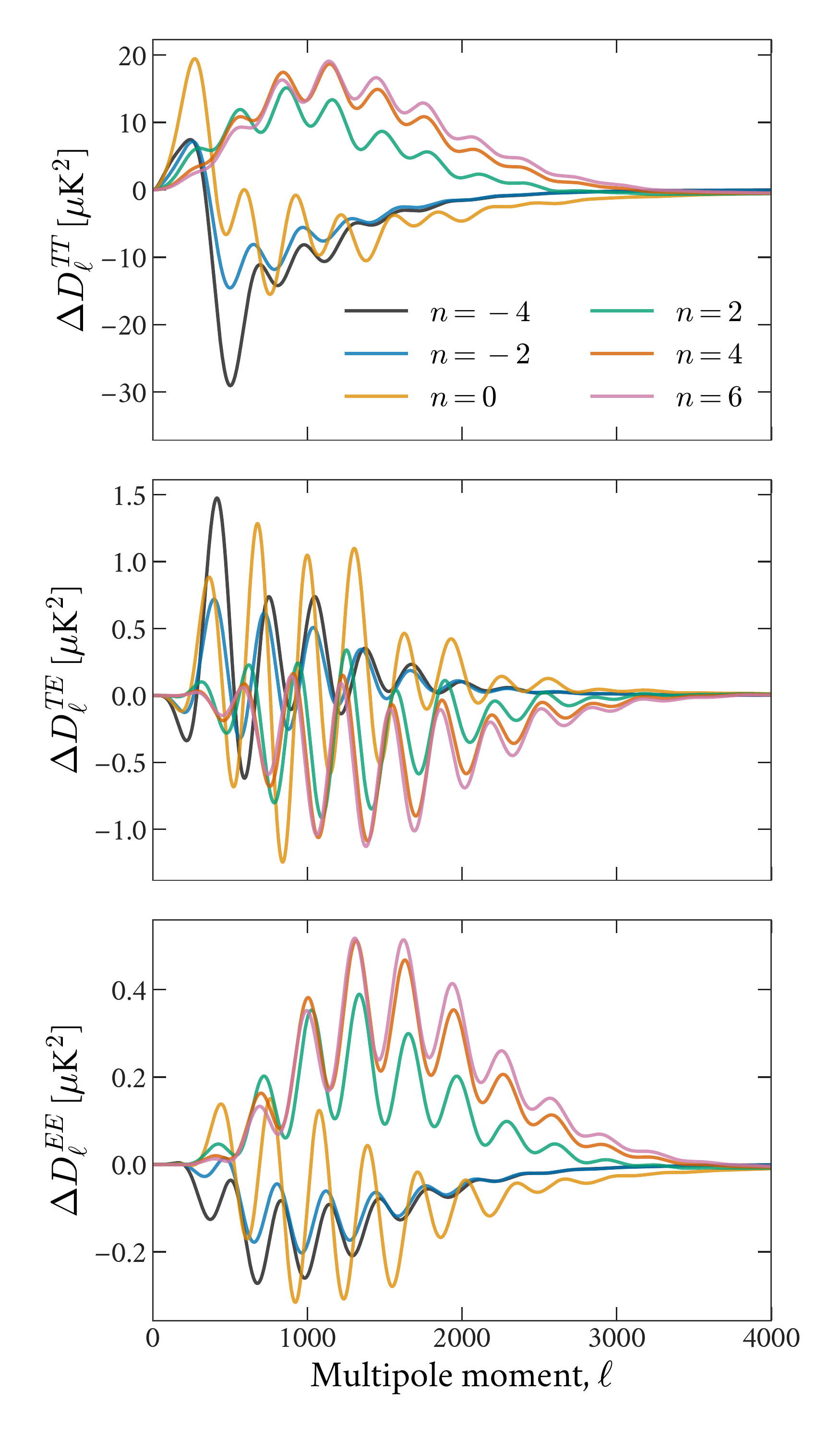}
\centering 
\caption{The effects on CMB anisotropy power spectra from DM-proton scattering for a variety of models described in Sec.~\ref{sec:models}. To demonstrate the overall small-scale suppression in power from the $n \geq 0$ models, we show the relative difference from a $\Lambda$CDM model for temperature correlations. To exhibit the effect on the acoustic peaks, we plot $D_{\ell} = \ell (\ell+1) C_{\ell} / 2 \pi$ for the temperature and polarisation auto- and cross-spectra. We show the case of $m_\chi=1$ GeV for scattering cross section with power law dependence on relative particle velocity, for power law indicies $n\in\{-4, -2, 0, 2, 4, 6\}$, setting the cross sections for visual clarity within an order of magnitude of their respective upper limits.
}
\label{fig:models}
\end{figure}

Scattering between dark matter and baryons (DM-baryon scattering) is an example of an extension of the standard model of cosmology that alters the dynamics of the early Universe, leaving fingerprints on the acoustic oscillations seen in the CMB \citep{chen2002, DvorkinBlum, BoddyGluscevic, GluscevicBoddy, Xu189710, 2021PhRvD.104j3521N,2022PhR...961....1B}.
Although a diverse set of astrophysical observations suggests that the cold dark matter in the Universe interacts only gravitationally, the possibility of a coupling between dark matter and baryons still presents a tantalizing target towards understanding the particle nature of dark matter. Direct detection experiments \citep[e.g.,][]{sensei2018, darkside2018, supercdms2019,  xenonnt2020} have made substantial progress in searching for such a coupling, typically through searching in the laboratory for nuclear recoils from scattering with dark matter.
Astrophysical constraints on dark matter can complement direct detection experiments, in particular delivering competitive sensitivities towards models with dark matter particles masses below approximately 1\,GeV for nuclear recoils and near to 1 MeV for electronic recoils. The most powerful astrophysical constraints on pre-recombination scattering between DM and baryons come from the Milky Way satellite abundance measurements \citep{Nadler:2020,2021PhRvL.126i1101N, 2021ApJ...907L..46M,2021PhRvD.104j3521N} and Lyman-$\alpha$-forest measurements \citep{2021arXiv211004024H, 2021arXiv211110386R}, while the CMB provides the most competitive bounds for post-recombination scattering \citep{SlatyerWu,2018PhRvD..98l3506B}. At the same time, constraints from the CMB data have very different sources of uncertainty than other observational probes, do not require modeling of baryonic physics arising from galaxy formation, and probe the same physics at different physical scales.
Notably, small-scale polarization measurements of the CMB have less contamination from astrophysical foregrounds relative to temperature anisotropies, and can carry valuable information about DM physics.

The features induced in CMB power spectra by scattering effects can vary, depending on how the momentum-transfer cross section $\sigma_{\mathrm{MT}}$ varies with relative velocity $v$; following the literature, we parameterize this dependence as a power law with index $n$, such that $\sigma_{\mathrm{MT}}= \sigma_0 \, v^n$ in natural units, for either scattering with protons, or electrons. In this work, the choice of $n$ amounts to the choice of the scattering model at hand.
DM-baryon scattering with power law index $n < 0$ tends to produce progressively stronger relative suppression of power in the CMB at small angular scales, as shown in Figure \ref{fig:models}, and represents scattering that takes place in the post-recombination universe (after $v$ redshifts due to the universal expansion), affecting the degree of lensing of the CMB. DM-baryon scattering with non-negative $n$ tends to also exhibit a substantial increase in power at intermediate scales ($\ell \sim 1000-2000$), by effectively increasing the mass of the baryons, and is dominated by scattering in the pre-recombination universe (where $v$ is driven by thermal velocities that are large at early times).
The \planck\, satellite measured the temperature anisotropies to the cosmic variance limit for scales up to $\ell \sim 2000$, but current and upcoming ground-based experiments such as the Atacama Cosmology Telescope \citep[ACT,][]{thornton/2016}, the South Pole Telescope \citep[SPT,][]{benson/etal:2014}, the Simons Observatory \citep[SO,][]{so_forecast:2019}, and CMB-S4 \citep{cmbs4:2019} promise to push the cosmic variance limit to $\ell \sim 3000 - 4000$ in both temperature and polarization.
As an example, the addition of ACT DR2 and SPT data improved the \planck{}-2015 constraint by a factor of two for some interaction models \citep{SlatyerWu}.

Previous CMB constraints were driven primarily by data from the \planck{} satellite \citep{DvorkinBlum, BoddyGluscevic, BoddyGluscevicPoulin, Xu189710, 2021PhRvD.104j3521N}. In this work, we use the ACT DR4 data \citep{choi2020, aiola2020}, collected during 2013$-$2016, to search for DM-baryon scattering.
In combination with the 2018 \planck{} data, we use these data to improve the CMB constraints on DM-baryon scattering.


In Sec.~\ref{sec:models}, we review how observables predicted by the Einstein-Boltzmann equations change in the presence of DM-proton scattering.
In Sec.~\ref{sec:data}, we describe the data we use for our analysis and detail the fitting procedure.
We present our main results in Sec.~\ref{sec:results} and conclude in Sec.~\ref{sec:conclusions}.

\section{Scattering model observables}
\label{sec:models}


Within the power-law parameterization for the momentum-transfer cross-section, the index $n=0$ arises in the simplest example of a spin-independent or spin-dependent contact interaction; millicharged DM exhibits Coulomb-like interaction with $n=-4$; $n=2$ corresponds to DM with an electric dipole moment \citep{Fitzpatrick:2010br}; 
In this study, we consider models with $n\in\{-4,-2,0,2,4,6\}$, as described in e.g., \cite{DvorkinBlum}. We separately constrain elastic scattering with protons and elastic scattering with electrons. 


Different values of $n$ lead to a different redshift evolution of the rate of momentum transfer $R_\chi$ between DM and baryons, affecting matter perturbations at different cosmological times: for $n=-4$, scattering is more important as thermal particle velocities decay, later on in cosmic history; for $n\geq-2$, scattering mainly occurs prior to recombination, at high redshift when thermal velocities are large \citep{BoddyGluscevic, BoddyGluscevicPoulin}.
However, all forms of scattering interactions considered here affect the matter distribution in the universe through collisional damping of small-scale perturbations. 
The resulting suppression of the matter transfer function is captured in the CMB temperature, polarization, and lensing power spectra. 
The main effect of the interactions is suppression of power at small angular scales. 
Secondary effects include small shifts in the acoustic peaks, as well as the increase in power on large angular scales.
The latter is particularly prominent in models where DM couples strongly to the baryon-photon fluid prior to recombination, producing an effective ``baryon-loading'' effect and increasing power at low multipoles \cite{BoddyGluscevic,BoddyGluscevicPoulin}. 

To accurately model the effects of DM-proton scattering on the CMB primary power spectra, we use a modified Boltzmann code CLASS \citep{class} developed for previous studies \citep{BoddyGluscevic} and publicly released with the work of \cite{2021PhRvD.104j3521N}.
This code includes scattering interactions and their effects on the matter transfer function and the thermal history.\footnote{\url{https://github.com/kboddy/class\_public/tree/dmeff}} 



\section{Data and methodology}
\label{sec:data}
\subsection{Cosmological Model} \label{sec:model}

We use Markov Chain Monte Carlo (MCMC) chains to sample the standard six parameters of the $\Lambda$CDM model, plus one or two extension parameters. The six $\Lambda$CDM parameters are
\begin{equation}
\left\{ n_s, \, \log \left( 10^{10}A_s \right), \, \tau_{\mathrm{reio}}, \, \Omega_b h^2, \, \Omega_c h^2, \, 100 \theta_s \right\}.
\end{equation}
for the scalar spectral index, scalar amplitude, optical depth to reionization, baryon density, cold dark matter density, and CMB peak position respectively.
The extension parameters are $\sigma_0$, the DM-baryon interaction cross section, and $m_{\chi}$, the DM particle mass. We treat each DM-baryon interaction cross section velocity dependence, $n$, individually as separate phenomenological models for analysis. For each choice of $n$, we perform one analysis in which scattering is limited only to protons, and a second with only electron scattering. 

To report confidence limits on cross section constraints, we sample the six $\Lambda$CDM parameters in addition to $\sigma_0$,
fixing $m_\chi$ at seven different values between 1 MeV and 1 TeV, and for $n\in\{-4,-2,0,2,4,6\}$. In these cases we impose a uniform prior on the cross section.
For exploration purposes we also estimate parameters for an eight-parameter model: $\Lambda$CDM plus $\log(\sigma_0)$ and $\log(m_\chi)$, in this case imposing no preference on the order of magnitude of the cross section and mass. Sampling both parameters simultaneously has not been done before in CMB analyses of DM-baryon scattering. For each mass, we choose a lower prior on $\log(\sigma_0)$ that is several decades below the lowest limit obtained from sampling $\sigma_0$ at fixed mass. This
choice of prior will have a small effect on numerical
results like the 95-percentile upper bound, as it removes a
small region of parameter space close to zero. We avoid
masses below 1 MeV for numerical stability.

We include an approximate treatment of neutrinos and other light relics by setting a massless light relic density of $N_{\mathrm{ur}} = 2.0328$ and including a single massive neutrino with mass $m_{\mathrm{ncdm}} = 0.06$ eV. In this work we replace all of the DM density in the universe with a component that interacts with baryons. However, we do retain a small tracer component of standard non-interacting cold dark matter (CDM) at the level of $10^{-12}$, in order to allow for numerical computation in the synchronous gauge of this component.

Existing formulae like {\tt halofit} \citep{halofit} are derived from N-body simulations which do not include DM-baryon scattering, and thus are unreliable for predicting effects of nonlinear growth on the late-time matter power spectrum in our extension cosmologies. We found {\tt halofit} produces unrealistic nonlinear matter power spectra when used in conjunction with dark matter scattering models, even for cross sections that result in only modest deviations from $\Lambda$CDM. Throughout this work, we include only linear $P(k)$ computations. Nonlinear growth tends to amplify power at small scales, which would tend to amplify CMB power spectra at scales most sensitive to dark matter scattering with baryons. Incorporating nonlinear growth thus amplifies the scattering signal relative to the instrumental noise. Thus, we argue that the bounds we present in this analysis are \emph{conservative} bounds derived from linear cosmology. This has particular impact on the lensing of the CMB, and may affect parameter constraints. We leave the treatment of nonlinear structure formation within dark matter-baryon interaction cosmologies for a later work, which we expect would provide even tighter constraints from the larger matter power signal. \cite{hill2021} showed that the ACT DR4 data were sufficiently precise to have a difference in cosmological parameters of order $0.2\sigma$ due to Boltzmann code precision settings; we also leave this implementation to future work. 

\subsection{Data and Sampling} \label{subsec:data}

In our main analysis, we use a combination of \planck{} 2018 and the Atacama Cosmology Telescope Data Release 4 (DR4). We use the foreground- and nuisance-marginalized versions of these likelihoods, representing the best estimates of the CMB bandpowers provided by these experiments. These marginalized likelihoods are Gaussian and each have one remaining nuisance parameter. We thus additionally sample over 
\begin{equation} \left\{ A_{\mathrm{Planck}}, \, y_p \right\}, \end{equation}
representing the \planck{} absolute calibration and ACT polarization efficiency respectively.

We perform this likelihood analysis using the sampling framework $\texttt{cobaya}$ \citep{cobaya}. We include $\texttt{planck\_2018\_highl\_plik.TTTEEE\_lite}$ and $\texttt{planck\_2018\_lowl.TT}$ for \planck{}, and $\texttt{pyactlike.ACTPol\_lite\_DR4}$ for ACT. We do not use the CMB lensing data from ACT. Following \cite{aiola2020}, we exclude $\ell < 1800$ data in $TT$ when combining the ACT data with \planck{}, in order to avoid double-counting the same sky measured at the cosmic variance limit. We also include a Gaussian prior on $\tau_{\mathrm{reio}} = 0.065 \pm 0.015$ to replace the large-scale polarization likelihood. We also experimented with the addition of some other common cosmological datasets (\planck{} low-$\ell$ polarization, \planck{} lensing, and BAO from SDSS DR12 \citep{sdss12}), but found these do not improve constraints on DM-baryon scattering.
The lack of improvement when including the \planck{} lensing is consistent with previous analyses with the \planck{} data \citep[e.g.,][]{BoddyGluscevic}, but we expect this to change with next-generation surveys \citep{LiGluscevic2018}.

\section{Results}

\label{sec:results}

\subsection{Velocity-Independent Constraints}

For the fiducial model of velocity-independent ($n=0$) scattering with a 1\,GeV DM particle, we find the inclusion of the ACT DR4 data reduces the upper limit on the cross section for proton scattering by $\sim 40$\%, with 95-percentile upper limits of
\begin{equation}
\sigma_0^{\text{GeV, }n=0} < \left\{
        \begin{array}{ll}
            4.7 \times 10^{-25} \,\mathrm{cm}^2 \quad \text{\textit{(Planck)}} \\
            2.9 \times 10^{-25} \,\mathrm{cm}^2 \quad \text{(\textit{Planck} + ACT DR4)}. 
        \end{array}
    \right.
\end{equation}
\begin{figure}[t]
\includegraphics[width=\columnwidth]{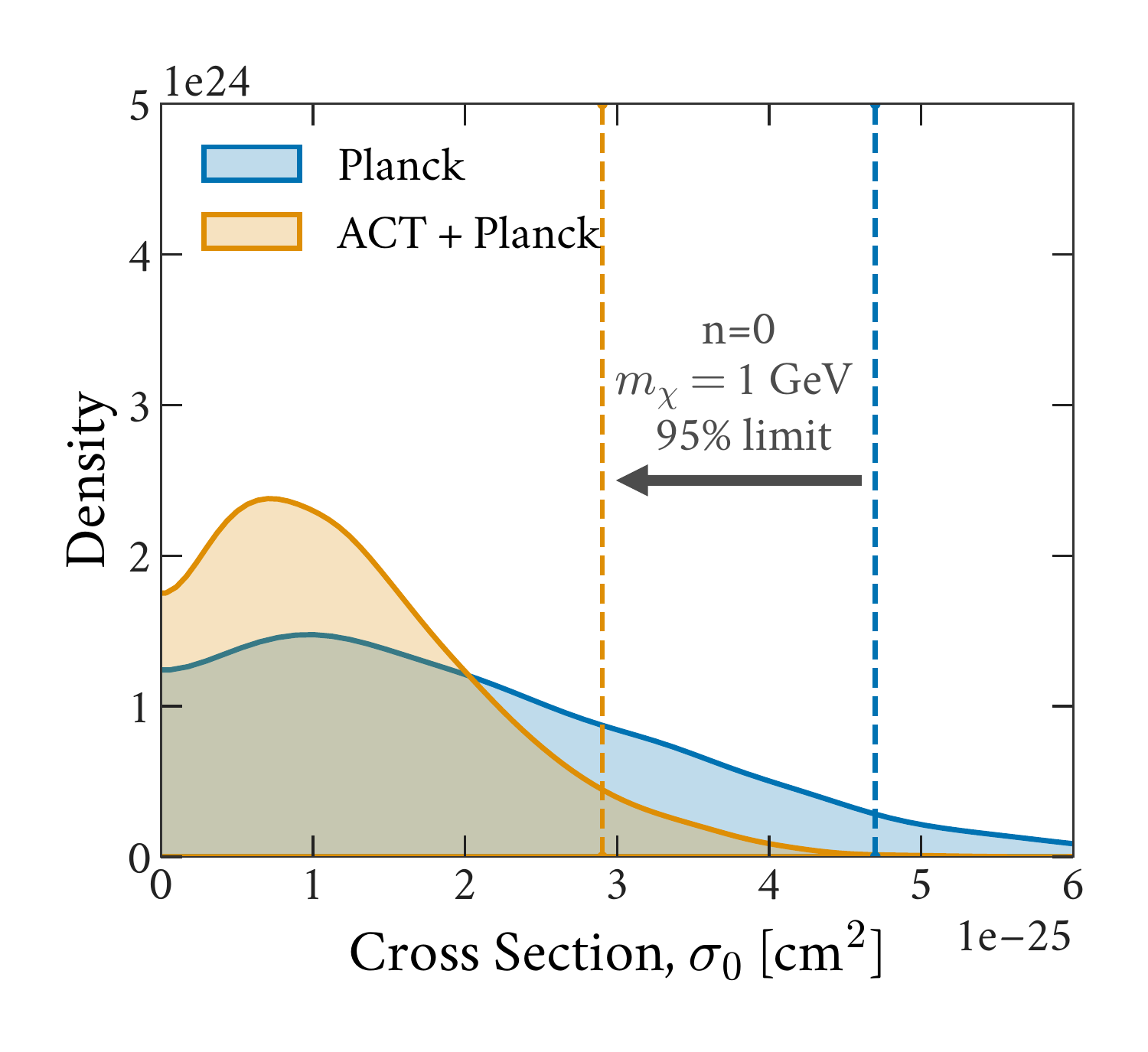}
\centering 
\caption{Marginalized 1D posteriors of the velocity-independent DM-proton scattering cross section, with the DM particle mass fixed at 1\,GeV. The dashed vertical lines show the 95\% upper limit from \planck{} data alone and the combination of \planck{} and ACT. The small-scale CMB data from ACT reduces the 95\% upper limits by $\sim 40$\% over constraints from \planck{} alone.}
\label{fig:marginalized1GeV}
\end{figure}
We illustrate these results in Figure \ref{fig:marginalized1GeV}, showing the marginalized 1D posterior of the DM scattering cross section. We find almost no correlation of this parameter with the $\Lambda$CDM parameters. For this model the data show no evidence for a nonzero cross section. This model demonstrates the constraining power of the ACT DR4 data, which provide improved measurements of the CMB damping tail and additional acoustic peaks in TE and EE, cutting the space of allowed cross sections compared to \planck{} alone. For other masses, and for the case of electron scattering, we provide upper limits derived from the \planck{} and ACT data in Appendix Table \ref{table:constraints}. 

Since the cross section parameter has a positive prior we check if the improved upper limit is compatible with expectation. In the Appendix we perform a Fisher matrix analysis, finding an expected $\sim$30\% improvement in errors from adding the ACT data to \planck{} for the $n=0$ model, consistent with our findings with the real data.

Both the \planck{} and ACT likelihoods used in this analysis include spectra and covariances that have been marginalized over models of foreground parameters. The effect of DM-proton and DM-electron scattering is imprinted in the CMB and is frequency-independent, but could still be biased by astrophysical foregrounds. ACT DR4 contains both additional small-scale information in temperature and polarization, but we expect the foreground contamination to primarily affect the temperature spectrum. The foregrounds in temperature primarily affect small-scale measurements, so we expect our analysis with the ACT DR4 temperature power spectra to be more susceptible to foreground contamination than previous work with lower resolution Planck data. However, we find that the $\Delta \chi^2$ arising from TT spectra at $\ell > 2000$ between the best-fit $\Lambda$CDM theory and DM-baryon scattering extension is less than half of the total $\Delta \chi^2$ arising from TT. 
We also confirm that the scattering cross section is not correlated with the \planck{} and ACT calibration nuisance parameters. 

\begin{figure*}[t]
  
\begin{tabular}{lll}
  \includegraphics[width=.3\linewidth]{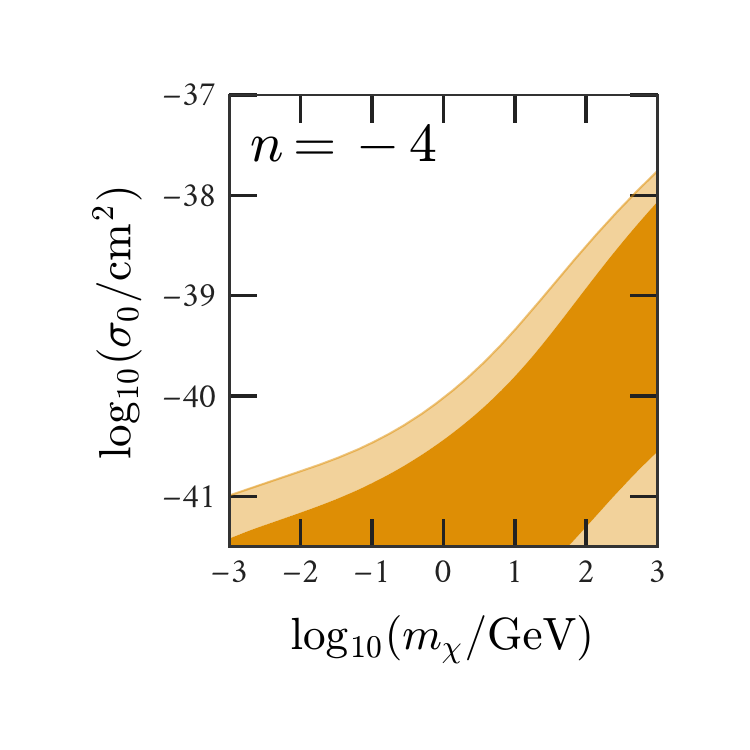} &
    \hspace*{-0.75cm} \includegraphics[width=.3\linewidth]{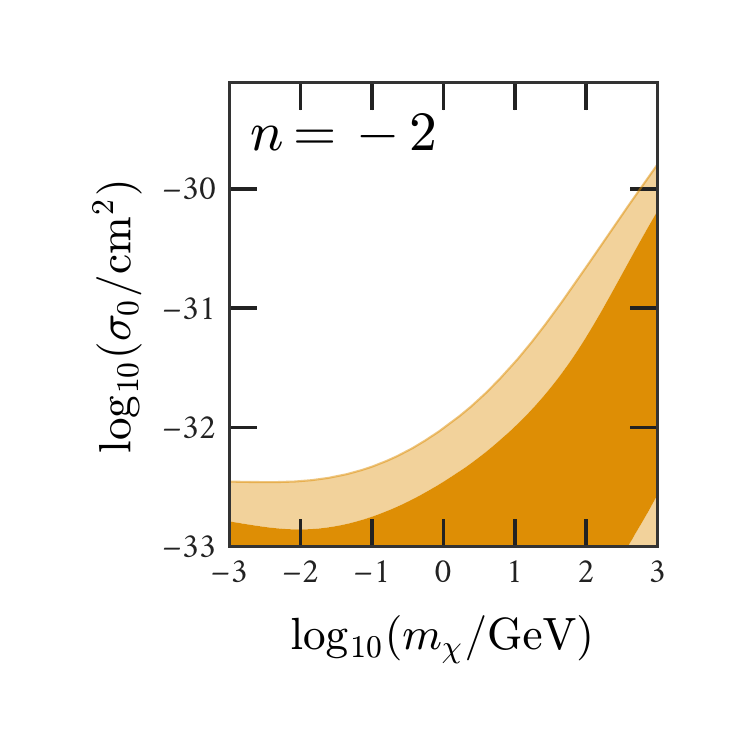} &
    \hspace*{-0.75cm} \includegraphics[width=.3\linewidth]{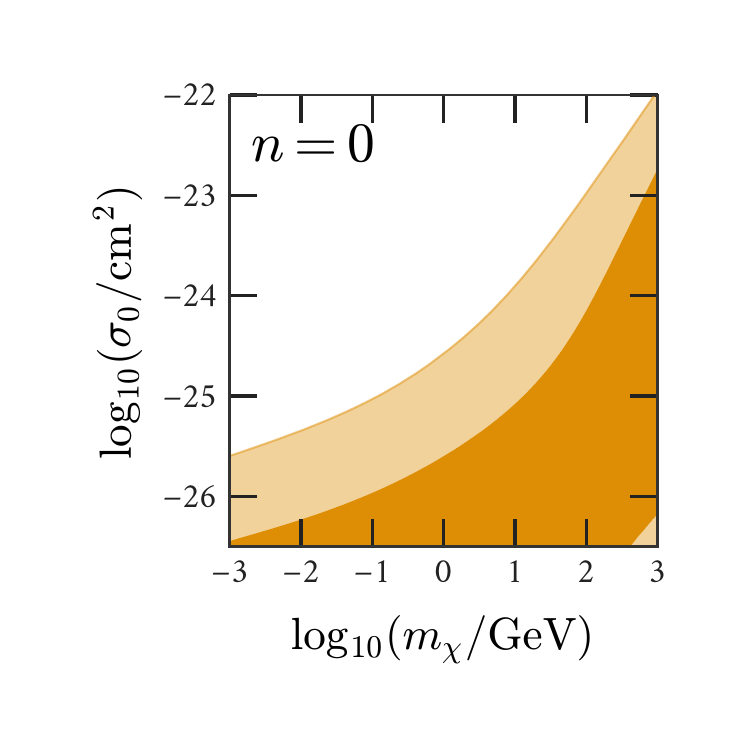}
   \vspace{-1.45cm}\\
  \includegraphics[width=.3\linewidth]{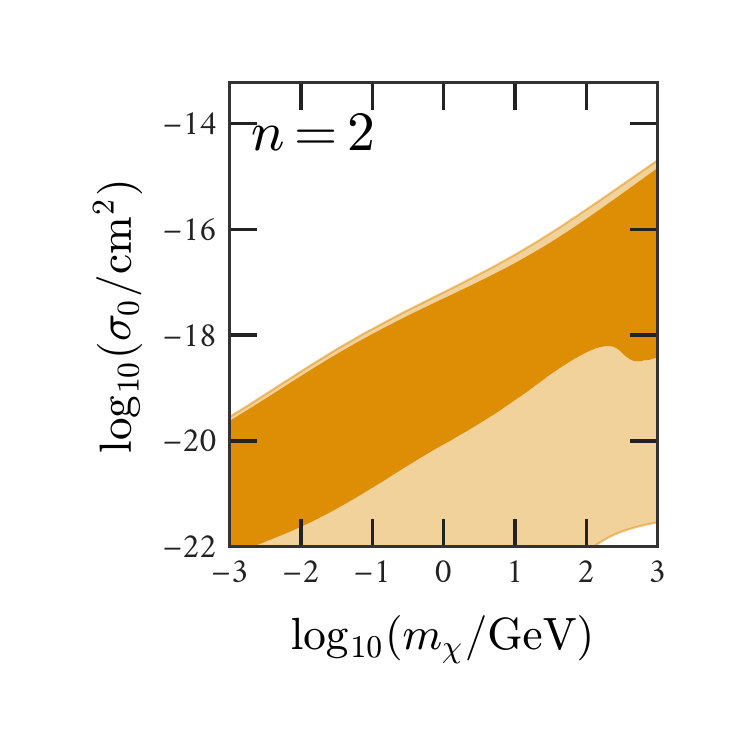} &
    \hspace*{-0.75cm} \includegraphics[width=.3\linewidth]{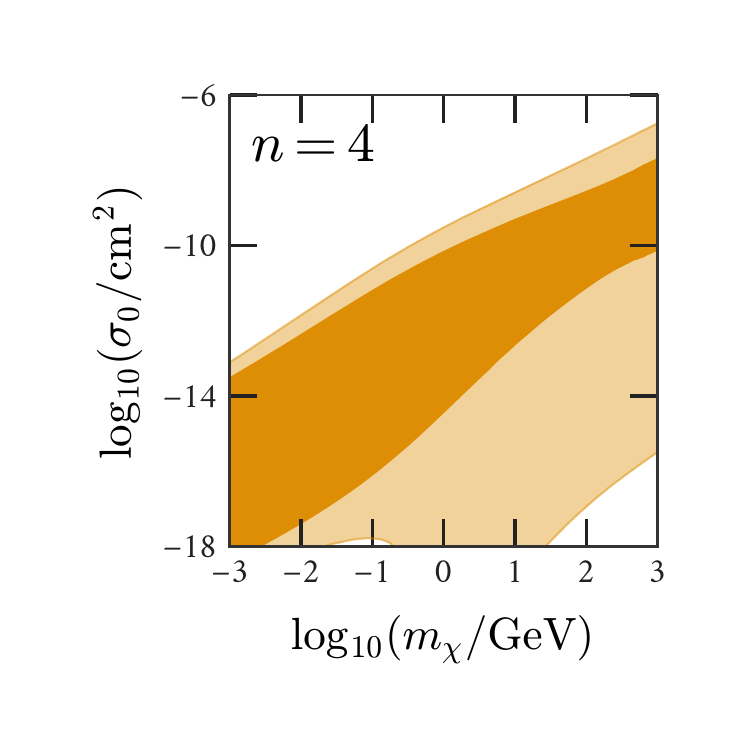} &
    \hspace*{-0.75cm} \includegraphics[width=.3\linewidth]{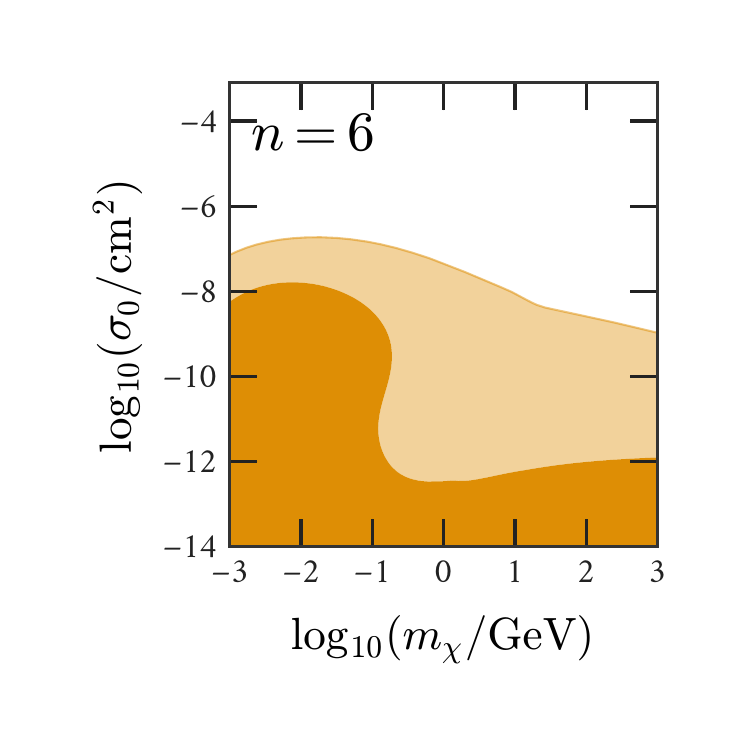} \\
    \vspace*{-0.75cm}
  \end{tabular}
\centering 
\caption{Posterior densities evaluated from the combined \planck{} and ACT likelihoods, sampled over all $\Lambda$CDM parameters jointly with the logarithms of the DM-proton cross section, $\sigma_0$, and dark matter particle mass, $m_\chi$. Contours shown are 68- and 95-percentile. We show all six proton-scattering models considered in this work ($n \in \{-4, -2, 0, 2, 4, 6\}$), sampling from 1 MeV to 1 TeV in dark matter particle mass.}
  \label{fig:freemass}
  \vspace{0.5cm}
\end{figure*}

In Figure \ref{fig:freemass} we show constraints from the eight-parameter model for proton scattering, where we simultaneously sample both $\log(m_{\chi})$ and $\log(\sigma_0)$, with $n=0$ shown in the upper right panel. We see a strong correlation between mass and the cross-section upper limit.


\subsection{Velocity-Dependent Constraints}
\label{subsec:veldepconstr}
\begin{figure}[t]
 \includegraphics[width=\columnwidth]{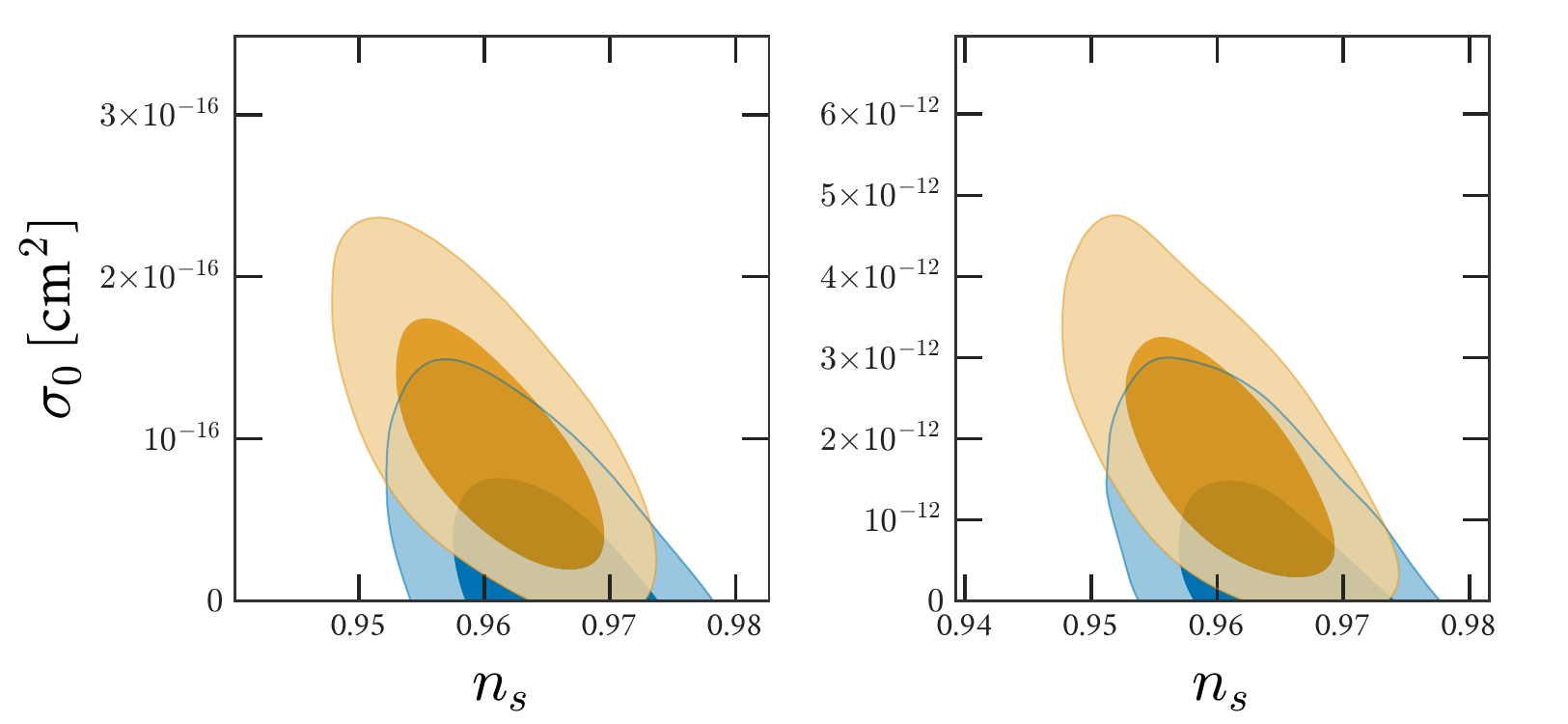} 
  \centering 
  \caption{The combined \planck{} and ACT DR4 data (orange), and \planck{} alone (blue), produce constraints on the DM-proton scattering cross section consistent with statistical fluctuations in a $\Lambda$CDM Universe. Contours shown are 68- and 95-percentile. We show the  posterior of the cross section for $n=2$ and $n=4$ models with a 1 MeV DM particle, together with $n_s$ which is the parameter most degenerate with the cross section. }
  \label{fig:linearMeV}
\end{figure}

We report results for the 7-parameter model ($\Lambda$CDM+$\sigma_0$) in Appendix Table \ref{table:constraints} for the suite of masses and model indices.\footnote{Our choice to report cross sections for a  linear rather than logarithmic prior can change constraints by up to a factor of two, which affects comparisons with previous work, e.g., \cite{GluscevicBoddy}.} When adding the ACT data we find posterior densities for proton scattering which have nonzero best-fitting scattering cross sections for $n=2$ and $n=4$. 
This is shown in Figure \ref{fig:linearMeV} for the 1 MeV case, for the cross section and the primordial tilt, $n_s$, which is most degenerate with the cross section. We find that the combined \planck{} and ACT posterior does not directly shrink inwards from the \planck{} constraints, but rather shifts upwards in cross section altogether by $\sim 1\sigma$.

The difference in goodness-of-fit for the ACT DR4 likelihood between a $\Lambda$CDM model and the best-fitting $n=2$, $m_{\chi}=1$ MeV model is driven primarily by the ACT temperature data, with
\begin{equation}
(\chi^2_{\Lambda\text{CDM}} - \chi^2_{n=2, \text{1 MeV}})_{\mathrm{ACT}} =  \left\{
        \begin{array}{llr}
            5.3 & & \text{\emph{(TT, TE, EE)}} \\
            4.0 & & \text{\emph{(TT)}} \\
            0.9 & & \text{\emph{(TE)}} \\
            -0.3 & & \text{\emph{(EE)}}.
        \end{array}
    \right.
\end{equation}
For two extra parameters applied to more than 100 degrees of freedom, this is not a significant improvement.
Overall from a model selection viewpoint we find no evidence for DM-baryon scattering in any of the six models considered in this work ($n \in \{-4, -2, 0, 2, 4, 6\}$). Approximately one in three datasets would randomly exhibit a similar $\sim 2 \sigma$ statistical fluctuation, when testing six models like $n \in \{-4, -2, 0, 2, 4, 6\}$, and for two additional parameters (mass and cross section). 

In testing the impact that ACT DR4 has on parameter constraints, we show in the Appendix that ACT improves on \textit{Planck} uncertainties by less than 10\% for $n<0$ models. For $n \ge 0$ models the improvement on the uncertainty is 30--50\%, with these models benefiting more from the smaller scale data. This also confirms that ACT would have been expected to improve constraints on $\sigma_0$ for these models, if not for a presumably statistical fluctuation towards nonzero best-fit values.


We show the derived posterior densities for the combined \planck{} and ACT data in Figure \ref{fig:freemass}, for the eight-parameter model for proton scattering. These samples are presented with a flat prior in the logarithm of the cross section, instead of the flat prior in the cross section used in \ref{fig:linearMeV}. The approach of sampling in the logarithms of the DM particle mass and cross section is useful for exploring the space of allowed models.
The CMB constraints exhibit a significant degeneracy between DM particle mass and cross section, and the allowed cross sections vary by several orders of magnitude as the mass changes from 1 MeV to 1 TeV. In the limit where the DM mass is much greater or much less than the mass of the scattering target, this degeneracy becomes a true power law. Although there are some masses for which the best-fitting cross-section is nonzero, 
we have no reason to believe that the DM particle mass takes on any particular value from 1 MeV to 1 TeV. CMB-only constraints struggle to break the degeneracy between DM particle mass and the scattering cross section. Bayesian analysis would marginalize over mass. Any such marginalization would erase a preference for nonzero cross section, as one can infer from Figure~\ref{fig:freemass}.

We also present constraints on DM-electron scattering in Table \ref{table:constraints}. These constraints exhibit behavior similar to the DM-proton scattering, with some modest 1$-$2$\sigma$ best-fit deviations from zero cross-section, but still consistent with $\Lambda$CDM. 



We test the effect of including additional \planck{} large-scale polarization in place of a prior on $\tau_{\text{reio}}$, as well as including \planck{} lensing and BAO constraints from SDSS DR12 \citep{sdss12}. 
The constraints on the DM-baryon interaction cross section are virtually unchanged with the inclusion of these data. There are the expected shifts for the optical depth to reionization, the amplitude $A_s$, and the DM density from these additional data sources, but these are not correlated with the DM-baryon scattering parameters.



\section{Conclusions and Discussion}
\label{sec:conclusions}

We have used new measurements of the CMB, particularly at small scales and in polarization, to look for evidence of elastic scattering between DM and baryons (protons and electrons). Compared to previous work, the inclusion of the ACT DR4 data provides more precise measurements of the high-$\ell$ acoustic peaks and damping tail in TE and EE. Relative to a $\Lambda$CDM model, the scattering models affect mostly the small scales, so the inclusion of the ACT DR4 dataset is especially suited to investigating this physics. Indeed, although the addition of the ACT DR4 likelihood  does not significantly improve constraints on the standard $\Lambda$CDM parameters \citep{aiola2020}, we find that for the fiducial model of velocity-independent dark matter scattering with a 1\,GeV dark matter particle, the combination of ACT DR4 and \planck{} improves the upper limit on the scattering cross section by $\sim 40\%$. 

The combined \planck{} and ACT likelihood yields posteriors consistent with statistical fluctuations about the non-scattering $\Lambda$CDM model, for all models considered in this work. However, many of the $n \neq 0$ models do exhibit a mild $< 2 \sigma$ deviation from zero cross section for many masses, as shown in Figures~\ref{fig:freemass} and \ref{fig:linearMeV}. This preference arises primarily from the ACT temperature data, but is not statistically significant. 

Since DM-baryon scattering reduces power at small scales, we expect that new high-resolution ground-based data, particularly measurements of the TE and EE correlations at high-$\ell$, will provide noteworthy improved constraints in the near future. Existing instruments like ACT and SPT, as well as future instruments like the Simons Observatory and CMB-S4, will provide as much as an order of magnitude in improvement for the scattering cross section. 

\begin{acknowledgments}
VG and RA acknowledge the support from NASA through the Astrophysics Theory Program, Award Number 21-ATP21-0135 and from the National Science Foundation under Grant No. PHY-2013951. KB acknowledges support from the NSF under Grant No.\ PHY-2112884. EC acknowledges support from the European Research Council (ERC) under the European Union’s Horizon 2020 research and innovation programme (Grant agreement No. 849169). JD is supported by NSF grants AST-1814971 and AST-2108126. ADH acknowledges support from the Sutton Family Chair in Science, Christianity and Cultures and from the Faculty of Arts and Science, University of Toronto. Research at Perimeter Institute is supported in part by the Government of Canada through the Department of Innovation, Science and Industry Canada and by the Province of Ontario through the Ministry of Colleges and Universities.
Computations were performed on Della as part of Princeton Research Computing resources at Princeton University.

Support for ACT was through the U.S.~National Science Foundation through awards AST-0408698, AST-0965625, and AST-1440226 for the ACT project, as well as awards PHY-0355328, PHY-0855887 and PHY-1214379. Funding was also provided by Princeton University, the University of Pennsylvania, and a Canada Foundation for Innovation (CFI) award to UBC.  ACT operates in the Parque Astron\'omico Atacama in northern Chile under the auspices of the Agencia Nacional de Investigaci\'on y Desarrollo (ANID).  The development of multichroic detectors and lenses was supported by NASA grants NNX13AE56G and NNX14AB58G.  Detector research at NIST was supported by the NIST Innovations in Measurement Science program. 
\end{acknowledgments}

\bibliographystyle{yahapj}
\bibliography{biblio.bib}

\begin{thebibliography}{}
\providecommand\natexlab[1]{#1}
\providecommand\JournalTitle[1]{#1}

\bibitem[{{Abazajian} {et~al.}(2019){Abazajian}, {Addison}, {Adshead}, {Ahmed},
  {Allen}, {Alonso}, {Alvarez}, {Anderson}, {Arnold}, {Baccigalupi}, {Bailey},
  {Barkats}, {Barron}, {Barry}, {Bartlett}, {Basu Thakur}, {Battaglia},
  {Baxter}, {Bean}, {Bebek}, {Bender}, {Benson}, {Berger}, {Bhimani},
  {Bischoff}, {Bleem}, {Bocquet}, {Boddy}, {Bonato}, {Bond}, {Borrill},
  {Bouchet}, {Brown}, {Bryan}, {Burkhart}, {Buza}, {Byrum}, {Calabrese},
  {Calafut}, {Caldwell}, {Carlstrom}, {Carron}, {Cecil}, {Challinor}, {Chang},
  {Chinone}, {Cho}, {Cooray}, {Crawford}, {Crites}, {Cukierman}, {Cyr-Racine},
  {de Haan}, {de Zotti}, {Delabrouille}, {Demarteau}, {Devlin}, {Di Valentino},
  {Dobbs}, {Duff}, {Duivenvoorden}, {Dvorkin}, {Edwards}, {Eimer}, {Errard},
  {Essinger-Hileman}, {Fabbian}, {Feng}, {Ferraro}, {Filippini}, {Flauger},
  {Flaugher}, {Fraisse}, {Frolov}, {Galitzki}, {Galli}, {Ganga}, {Gerbino},
  {Gilchriese}, {Gluscevic}, {Green}, {Grin}, {Grohs}, {Gualtieri}, {Guarino},
  {Gudmundsson}, {Habib}, {Haller}, {Halpern}, {Halverson}, {Hanany},
  {Harrington}, {Hasegawa}, {Hasselfield}, {Hazumi}, {Heitmann}, {Henderson},
  {Henning}, {Hill}, {Hlozek}, {Holder}, {Holzapfel}, {Hubmayr},
  {Huffenberger}, {Huffer}, {Hui}, {Irwin}, {Johnson}, {Johnstone}, {Jones},
  {Karkare}, {Katayama}, {Kerby}, {Kernovsky}, {Keskitalo}, {Kisner}, {Knox},
  {Kosowsky}, {Kovac}, {Kovetz}, {Kuhlmann}, {Kuo}, {Kurita}, {Kusaka},
  {Lahteenmaki}, {Lawrence}, {Lee}, {Lewis}, {Li}, {Linder}, {Loverde},
  {Lowitz}, {Madhavacheril}, {Mantz}, {Matsuda}, {Mauskopf}, {McMahon},
  {McQuinn}, {Meerburg}, {Melin}, {Meyers}, {Millea}, {Mohr}, {Moncelsi},
  {Mroczkowski}, {Mukherjee}, {M{\"u}nchmeyer}, {Nagai}, {Nagy}, {Namikawa},
  {Nati}, {Natoli}, {Negrello}, {Newburgh}, {Niemack}, {Nishino}, {Nordby},
  {Novosad}, {O'Connor}, {Obied}, {Padin}, {Pandey}, {Partridge}, {Pierpaoli},
  {Pogosian}, {Pryke}, {Puglisi}, {Racine}, {Raghunathan}, {Rahlin},
  {Rajagopalan}, {Raveri}, {Reichanadter}, {Reichardt}, {Remazeilles}, {Rocha},
  {Roe}, {Roy}, {Ruhl}, {Salatino}, {Saliwanchik}, {Schaan}, {Schillaci},
  {Schmittfull}, {Scott}, {Sehgal}, {Shandera}, {Sheehy}, {Sherwin},
  {Shirokoff}, {Simon}, {Slosar}, {Somerville}, {Spergel}, {Staggs}, {Stark},
  {Stompor}, {Story}, {Stoughton}, {Suzuki}, {Tajima}, {Teply}, {Thompson},
  {Timbie}, {Tomasi}, {Treu}, {Tristram}, {Tucker}, {Umilt{\`a}}, {van
  Engelen}, {Vieira}, {Vieregg}, {Vogelsberger}, {Wang}, {Watson}, {White},
  {Whitehorn}, {Wollack}, {Kimmy Wu}, {Xu}, {Yasini}, {Yeck}, {Yoon}, {Young},
  \& {Zonca}}]{cmbs4:2019}
{Abazajian}, K., {Addison}, G., {Adshead}, P., {et~al.} 2019,
  \JournalTitle{arXiv e-prints}, arXiv:1907.04473

\bibitem[{{Agnes} {et~al.}(2018){Agnes}, {Albuquerque}, {Alexander}, {Alton},
  {Araujo}, {Asner}, {Ave}, {Back}, {Baldin}, {Batignani}, {Biery}, {Bocci},
  {Bonfini}, {Bonivento}, {Bottino}, {Budano}, {Bussino}, {Cadeddu}, {Cadoni},
  {Calaprice}, {Caminata}, {Canci}, {Candela}, {Caravati}, {Cariello},
  {Carlini}, {Carpinelli}, {Catalanotti}, {Cataudella}, {Cavalcante},
  {Cavuoti}, {Cereseto}, {Chepurnov}, {Cical{\`o}}, {Cifarelli}, {Cocco},
  {Covone}, {D'Angelo}, {D'Incecco}, {D'Urso}, {Davini}, {De Candia}, {De
  Cecco}, {De Deo}, {De Filippis}, {De Rosa}, {De Vincenzi}, {Demontis},
  {Derbin}, {Devoto}, {Di Eusanio}, {Di Pietro}, {Dionisi}, {Downing},
  {Edkins}, {Empl}, {Fan}, {Fiorillo}, {Fomenko}, {Franco}, {Gabriele},
  {Gabrieli}, {Galbiati}, {Garcia Abia}, {Ghiano}, {Giagu}, {Giganti},
  {Giovanetti}, {Gorchakov}, {Goretti}, {Granato}, {Gromov}, {Guan},
  {Guardincerri}, {Gulino}, {Hackett}, {Hassanshahi}, {Herner}, {Hosseini},
  {Hughes}, {Humble}, {Hungerford}, {Ianni}, {Ianni}, {Ippolito}, {James},
  {Johnson}, {Kahn}, {Keeter}, {Kendziora}, {Kochanek}, {Koh}, {Korablev},
  {Korga}, {Kubankin}, {Kuss}, {La Commara}, {Lai}, {Li}, {Lisanti}, {Lissia},
  {Loer}, {Longo}, {Ma}, {Machado}, {Machulin}, {Mandarano}, {Mapelli}, {Mari},
  {Maricic}, {Martoff}, {Messina}, {Meyers}, {Milincic}, {Mishra-Sharma},
  {Monte}, {Morrocchi}, {Mount}, {Muratova}, {Musico}, {Nania}, {Navrer
  Agasson}, {Nozdrina}, {Oleinik}, {Orsini}, {Ortica}, {Pagani}, {Pallavicini},
  {Pandola}, {Pantic}, {Paoloni}, {Pazzona}, {Pelczar}, {Pelliccia}, {Pesudo},
  {Picciau}, {Pocar}, {Pordes}, {Poudel}, {Pugachev}, {Qian}, {Ragusa},
  {Razeti}, {Razeto}, {Reinhold}, {Renshaw}, {Rescigno}, {Riffard}, {Romani},
  {Rossi}, {Rossi}, {Sablone}, {Samoylov}, {Sands}, {Sanfilippo}, {Sant},
  {Santorelli}, {Savarese}, {Scapparone}, {Schlitzer}, {Segreto}, {Semenov},
  {Shchagin}, {Sheshukov}, {Singh}, {Skorokhvatov}, {Smirnov}, {Sotnikov},
  {Stanford}, {Stracka}, {Suffritti}, {Suvorov}, {Tartaglia}, {Testera},
  {Tonazzo}, {Trinchese}, {Unzhakov}, {Verducci}, {Vishneva}, {Vogelaar},
  {Wada}, {Waldrop}, {Wang}, {Wang}, {Watson}, {Westerdale}, {Wojcik},
  {Wojcik}, {Xiang}, {Xiao}, {Yang}, {Ye}, {Zhu}, {Zichichi}, {Zuzel}, \&
  {DarkSide Collaboration}}]{darkside2018}
{Agnes}, P., {Albuquerque}, I.~F.~M., {Alexander}, T., {et~al.} 2018,
  \href{http://dx.doi.org/10.1103/PhysRevLett.121.111303}{\JournalTitle{\prl},
  121, 111303}

\bibitem[{{Agnese} {et~al.}(2019){Agnese}, {Aralis}, {Aramaki}, {Arnquist},
  {Azadbakht}, {Baker}, {Banik}, {Barker}, {Bauer}, {Binder}, {Bowles},
  {Brink}, {Bunker}, {Cabrera}, {Calkins}, {Cameron}, {Cartaro}, {Cerde{\~n}o},
  {Chang}, {Cooley}, {Cornell}, {Cushman}, {De Brienne}, {Doughty}, {Fascione},
  {Figueroa-Feliciano}, {Fink}, {Fritts}, {Gerbier}, {Germond}, {Ghaith},
  {Golwala}, {Harris}, {Herbert}, {Hong}, {Hoppe}, {Hsu}, {Huber}, {Iyer},
  {Jardin}, {Jastram}, {Jena}, {Kelsey}, {Kennedy}, {Kubik}, {Kurinsky},
  {Lawrence}, {Loer}, {Lopez Asamar}, {Lukens}, {MacDonell}, {Mahapatra},
  {Mandic}, {Mast}, {Miller}, {Mirabolfathi}, {Mohanty}, {Morales Mendoza},
  {Nelson}, {Neog}, {Orrell}, {Oser}, {Page}, {Partridge}, {Pepin}, {Ponce},
  {Poudel}, {Pyle}, {Qiu}, {Rau}, {Reisetter}, {Ren}, {Reynolds}, {Roberts},
  {Robinson}, {Rogers}, {Saab}, {Sadoulet}, {Sander}, {Scarff}, {Schnee},
  {Scorza}, {Senapati}, {Serfass}, {Speller}, {Stanford}, {Stein}, {Street},
  {Tanaka}, {Toback}, {Underwood}, {Villano}, {von Krosigk}, {Watkins},
  {Wilson}, {Wilson}, {Winchell}, {Wright}, {Yellin}, {Young}, {Zhang}, {Zhao},
  \& {SuperCDMS Collaboration}}]{supercdms2019}
{Agnese}, R., {Aralis}, T., {Aramaki}, T., {et~al.} 2019,
  \href{http://dx.doi.org/10.1103/PhysRevD.99.062001}{\JournalTitle{\prd}, 99,
  062001}

\bibitem[{{Aiola} {et~al.}(2020){Aiola}, {Calabrese}, {Maurin}, {Naess},
  {Schmitt}, {Abitbol}, {Addison}, {Ade}, {Alonso}, {Amiri}, {Amodeo},
  {Angile}, {Austermann}, {Baildon}, {Battaglia}, {Beall}, {Bean}, {Becker},
  {Bond}, {Bruno}, {Calafut}, {Campusano}, {Carrero}, {Chesmore}, {Cho},
  {Choi}, {Clark}, {Cothard}, {Crichton}, {Crowley}, {Darwish}, {Datta},
  {Denison}, {Devlin}, {Duell}, {Duff}, {Duivenvoorden}, {Dunkley},
  {D{\"u}nner}, {Essinger-Hileman}, {Fankhanel}, {Ferraro}, {Fox}, {Fuzia},
  {Gallardo}, {Gluscevic}, {Golec}, {Grace}, {Gralla}, {Guan}, {Hall},
  {Halpern}, {Han}, {Hargrave}, {Hasselfield}, {Helton}, {Henderson},
  {Hensley}, {Hill}, {Hilton}, {Hilton}, {Hincks}, {Hlo{\v{z}}ek}, {Ho},
  {Hubmayr}, {Huffenberger}, {Hughes}, {Infante}, {Irwin}, {Jackson}, {Klein},
  {Knowles}, {Koopman}, {Kosowsky}, {Lakey}, {Li}, {Li}, {Li}, {Lokken},
  {Louis}, {Lungu}, {MacInnis}, {Madhavacheril}, {Maldonado}, {Mallaby-Kay},
  {Marsden}, {McMahon}, {Menanteau}, {Moodley}, {Morton}, {Namikawa}, {Nati},
  {Newburgh}, {Nibarger}, {Nicola}, {Niemack}, {Nolta}, {Orlowski-Sherer},
  {Page}, {Pappas}, {Partridge}, {Phakathi}, {Pisano}, {Prince}, {Puddu}, {Qu},
  {Rivera}, {Robertson}, {Rojas}, {Salatino}, {Schaan}, {Schillaci}, {Sehgal},
  {Sherwin}, {Sierra}, {Sievers}, {Sifon}, {Sikhosana}, {Simon}, {Spergel},
  {Staggs}, {Stevens}, {Storer}, {Sunder}, {Switzer}, {Thorne}, {Thornton},
  {Trac}, {Treu}, {Tucker}, {Vale}, {Van Engelen}, {Van Lanen}, {Vavagiakis},
  {Wagoner}, {Wang}, {Ward}, {Wollack}, {Xu}, {Zago}, \& {Zhu}}]{aiola2020}
{Aiola}, S., {Calabrese}, E., {Maurin}, L., {et~al.} 2020,
  \href{http://dx.doi.org/10.1088/1475-7516/2020/12/047}{\JournalTitle{\jcap},
  2020, 047}

\bibitem[{{Alam} {et~al.}(2017){Alam}, {Ata}, {Bailey}, {Beutler}, {Bizyaev},
  {Blazek}, {Bolton}, {Brownstein}, {Burden}, {Chuang}, {Comparat}, {Cuesta},
  {Dawson}, {Eisenstein}, {Escoffier}, {Gil-Mar{\'\i}n}, {Grieb}, {Hand}, {Ho},
  {Kinemuchi}, {Kirkby}, {Kitaura}, {Malanushenko}, {Malanushenko}, {Maraston},
  {McBride}, {Nichol}, {Olmstead}, {Oravetz}, {Padmanabhan},
  {Palanque-Delabrouille}, {Pan}, {Pellejero-Ibanez}, {Percival}, {Petitjean},
  {Prada}, {Price-Whelan}, {Reid}, {Rodr{\'\i}guez-Torres}, {Roe}, {Ross},
  {Ross}, {Rossi}, {Rubi{\~n}o-Mart{\'\i}n}, {Saito}, {Salazar-Albornoz},
  {Samushia}, {S{\'a}nchez}, {Satpathy}, {Schlegel}, {Schneider},
  {Sc{\'o}ccola}, {Seo}, {Sheldon}, {Simmons}, {Slosar}, {Strauss}, {Swanson},
  {Thomas}, {Tinker}, {Tojeiro}, {Maga{\~n}a}, {Vazquez}, {Verde}, {Wake},
  {Wang}, {Weinberg}, {White}, {Wood-Vasey}, {Y{\`e}che}, {Zehavi}, {Zhai}, \&
  {Zhao}}]{sdss12}
{Alam}, S., {Ata}, M., {Bailey}, S., {et~al.} 2017,
  \href{http://dx.doi.org/10.1093/mnras/stx721}{\JournalTitle{\mnras}, 470,
  2617}

\bibitem[{{Aprile} {et~al.}(2020){Aprile}, {Aalbers}, {Agostini}, {Alfonsi},
  {Althueser}, {Amaro}, {Antochi}, {Angelino}, {Angevaare}, {Arneodo}, {Barge},
  {Baudis}, {Bauermeister}, {Bellagamba}, {Benabderrahmane}, {Berger}, {Brown},
  {Brown}, {Bruenner}, {Bruno}, {Budnik}, {Capelli}, {Cardoso}, {Cichon},
  {Cimmino}, {Clark}, {Coderre}, {Colijn}, {Conrad}, {Cussonneau}, {Decowski},
  {Depoian}, {Di Gangi}, {Di Giovanni}, {Di Stefano}, {Diglio}, {Elykov},
  {Eurin}, {Ferella}, {Fulgione}, {Gaemers}, {Gaior}, {Galloway}, {Gao},
  {Grandi}, {Hasterok}, {Hils}, {Hiraide}, {Hoetzsch}, {Howlett}, {Iacovacci},
  {Itow}, {Joerg}, {Kato}, {Kazama}, {Kobayashi}, {Koltman}, {Kopec},
  {Landsman}, {Lang}, {Levinson}, {Lin}, {Lindemann}, {Lindner}, {Lombardi},
  {Long}, {Lopes}, {L{\'o}pez Fune}, {Macolino}, {Mahlstedt}, {Mancuso},
  {Manenti}, {Manfredini}, {Marignetti}, {Marrod{\'a}n Undagoitia}, {Martens},
  {Masbou}, {Masson}, {Mastroianni}, {Messina}, {Miuchi}, {Mizukoshi},
  {Molinario}, {Mor{\r{a}}}, {Moriyama}, {Mosbacher}, {Murra}, {Naganoma},
  {Ni}, {Oberlack}, {Odgers}, {Palacio}, {Pelssers}, {Peres}, {Pienaar},
  {Pizzella}, {Plante}, {Qin}, {Qiu}, {Ram{\'\i}rez Garc{\'\i}a}, {Reichard},
  {Rocchetti}, {Rupp}, {dos Santos}, {Sartorelli}, {{\v{S}}ar{\v{c}}evi{\'c}},
  {Scheibelhut}, {Schreiner}, {Schulte}, {Schumann}, {Scotto Lavina}, {Selvi},
  {Semeria}, {Shagin}, {Shockley}, {Silva}, {Simgen}, {Takeda}, {Therreau},
  {Thers}, {Toschi}, {Trinchero}, {Tunnell}, {Valerius}, {Vargas}, {Volta},
  {Wang}, {Wei}, {Weinheimer}, {Weiss}, {Wenz}, {Wittweg}, {Xu}, {Yamashita},
  {Ye}, {Zavattini}, {Zhang}, {Zhu}, \& {Zopounidis}}]{xenonnt2020}
{Aprile}, E., {Aalbers}, J., {Agostini}, F., {et~al.} 2020,
  \href{http://dx.doi.org/10.1088/1475-7516/2020/11/031}{\JournalTitle{\jcap},
  2020, 031}

\bibitem[{{Benson} {et~al.}(2014){Benson}, {Ade}, {Ahmed}, {Allen}, {Arnold},
  {Austermann}, {Bender}, {Bleem}, {Carlstrom}, {Chang}, {Cho}, {Cliche},
  {Crawford}, {Cukierman}, {de Haan}, {Dobbs}, {Dutcher}, {Everett}, {Gilbert},
  {Halverson}, {Hanson}, {Harrington}, {Hattori}, {Henning}, {Hilton},
  {Holder}, {Holzapfel}, {Irwin}, {Keisler}, {Knox}, {Kubik}, {Kuo}, {Lee},
  {Leitch}, {Li}, {McDonald}, {Meyer}, {Montgomery}, {Myers}, {Natoli},
  {Nguyen}, {Novosad}, {Padin}, {Pan}, {Pearson}, {Reichardt}, {Ruhl},
  {Saliwanchik}, {Simard}, {Smecher}, {Sayre}, {Shirokoff}, {Stark}, {Story},
  {Suzuki}, {Thompson}, {Tucker}, {Vanderlinde}, {Vieira}, {Vikhlinin}, {Wang},
  {Yefremenko}, \& {Yoon}}]{benson/etal:2014}
{Benson}, B.~A., {Ade}, P.~A.~R., {Ahmed}, Z., {et~al.} 2014,
  \href{http://dx.doi.org/10.1117/12.2057305}{in Society of Photo-Optical
  Instrumentation Engineers (SPIE) Conference Series, Vol. 9153, Millimeter,
  Submillimeter, and Far-Infrared Detectors and Instrumentation for Astronomy
  VII, ed. W.~S. {Holland} \& J.~{Zmuidzinas}}, 91531P

\bibitem[{Bezanson {et~al.}(2017)Bezanson, Edelman, Karpinski, \&
  Shah}]{bezanson2017julia}
Bezanson, J., Edelman, A., Karpinski, S., \& Shah, V.~B. 2017,
  \href{https://doi.org/10.1137/141000671}{\JournalTitle{SIAM review}, 59, 65}

\bibitem[{{Boddy} \& {Gluscevic}(2018)}]{BoddyGluscevic}
{Boddy}, K.~K., \& {Gluscevic}, V. 2018,
  \href{http://dx.doi.org/10.1103/PhysRevD.98.083510}{\JournalTitle{\prd}, 98,
  083510}

\bibitem[{{Boddy} {et~al.}(2018{\natexlab{a}}){Boddy}, {Gluscevic}, {Poulin},
  {Kovetz}, {Kamionkowski}, \& {Barkana}}]{2018PhRvD..98l3506B}
{Boddy}, K.~K., {Gluscevic}, V., {Poulin}, V., {et~al.} 2018{\natexlab{a}},
  \href{http://dx.doi.org/10.1103/PhysRevD.98.123506}{\JournalTitle{\prd}, 98,
  123506}

\bibitem[{{Boddy} {et~al.}(2018{\natexlab{b}}){Boddy}, {Gluscevic}, {Poulin},
  {Kovetz}, {Kamionkowski}, \& {Barkana}}]{BoddyGluscevicPoulin}
---. 2018{\natexlab{b}},
  \href{http://dx.doi.org/10.1103/PhysRevD.98.123506}{\JournalTitle{\prd}, 98,
  123506}

\bibitem[{{Boyarsky} {et~al.}(2009){Boyarsky}, {Ruchayskiy}, \&
  {Shaposhnikov}}]{boyarsky2009}
{Boyarsky}, A., {Ruchayskiy}, O., \& {Shaposhnikov}, M. 2009,
  \href{http://dx.doi.org/10.1146/annurev.nucl.010909.083654}{\JournalTitle{Annual
  Review of Nuclear and Particle Science}, 59, 191}

\bibitem[{{Brinckmann} {et~al.}(2019){Brinckmann}, {Hooper}, {Archidiacono},
  {Lesgourgues}, \& {Sprenger}}]{brinckmann2019}
{Brinckmann}, T., {Hooper}, D.~C., {Archidiacono}, M., {Lesgourgues}, J., \&
  {Sprenger}, T. 2019,
  \href{http://dx.doi.org/10.1088/1475-7516/2019/01/059}{\JournalTitle{\jcap},
  2019, 059}

\bibitem[{{Brust} {et~al.}(2013){Brust}, {Kaplan}, \& {Walters}}]{brust2013}
{Brust}, C., {Kaplan}, D.~E., \& {Walters}, M.~T. 2013,
  \href{http://dx.doi.org/10.1007/JHEP12(2013)058}{\JournalTitle{Journal of
  High Energy Physics}, 2013, 58}

\bibitem[{{Buen-Abad} {et~al.}(2022){Buen-Abad}, {Essig}, {McKeen}, \&
  {Zhong}}]{2022PhR...961....1B}
{Buen-Abad}, M.~A., {Essig}, R., {McKeen}, D., \& {Zhong}, Y.-M. 2022,
  \href{http://dx.doi.org/10.1016/j.physrep.2022.02.006}{\JournalTitle{\physrep},
  961, 1}

\bibitem[{{Buen-Abad} {et~al.}(2018){Buen-Abad}, {Schmaltz}, {Lesgourgues}, \&
  {Brinckmann}}]{buenabad2018}
{Buen-Abad}, M.~A., {Schmaltz}, M., {Lesgourgues}, J., \& {Brinckmann}, T.
  2018,
  \href{http://dx.doi.org/10.1088/1475-7516/2018/01/008}{\JournalTitle{\jcap},
  2018, 008}

\bibitem[{{Chen} {et~al.}(2002){Chen}, {Hannestad}, \& {Scherrer}}]{chen2002}
{Chen}, X., {Hannestad}, S., \& {Scherrer}, R.~J. 2002, \JournalTitle{arXiv
  e-prints}, {arXiv astro ph}/0202496

\bibitem[{{Choi} {et~al.}(2020){Choi}, {Hasselfield}, {Ho}, {Koopman}, {Lungu},
  {Abitbol}, {Addison}, {Ade}, {Aiola}, {Alonso}, {Amiri}, {Amodeo}, {Angile},
  {Austermann}, {Baildon}, {Battaglia}, {Beall}, {Bean}, {Becker}, {Bond},
  {Bruno}, {Calabrese}, {Calafut}, {Campusano}, {Carrero}, {Chesmore}, {Cho},
  {Clark}, {Cothard}, {Crichton}, {Crowley}, {Darwish}, {Datta}, {Denison},
  {Devlin}, {Duell}, {Duff}, {Duivenvoorden}, {Dunkley}, {D{\"u}nner},
  {Essinger-Hileman}, {Fankhanel}, {Ferraro}, {Fox}, {Fuzia}, {Gallardo},
  {Gluscevic}, {Golec}, {Grace}, {Gralla}, {Guan}, {Hall}, {Halpern}, {Han},
  {Hargrave}, {Henderson}, {Hensley}, {Hill}, {Hilton}, {Hilton}, {Hincks},
  {Hlo{\v{z}}ek}, {Hubmayr}, {Huffenberger}, {Hughes}, {Infante}, {Irwin},
  {Jackson}, {Klein}, {Knowles}, {Kosowsky}, {Lakey}, {Li}, {Li}, {Li},
  {Lokken}, {Louis}, {MacInnis}, {Madhavacheril}, {Maldonado}, {Mallaby-Kay},
  {Marsden}, {Maurin}, {McMahon}, {Menanteau}, {Moodley}, {Morton}, {Naess},
  {Namikawa}, {Nati}, {Newburgh}, {Nibarger}, {Nicola}, {Niemack}, {Nolta},
  {Orlowski-Sherer}, {Page}, {Pappas}, {Partridge}, {Phakathi}, {Prince},
  {Puddu}, {Qu}, {Rivera}, {Robertson}, {Rojas}, {Salatino}, {Schaan},
  {Schillaci}, {Schmitt}, {Sehgal}, {Sherwin}, {Sierra}, {Sievers}, {Sifon},
  {Sikhosana}, {Simon}, {Spergel}, {Staggs}, {Stevens}, {Storer}, {Sunder},
  {Switzer}, {Thorne}, {Thornton}, {Trac}, {Treu}, {Tucker}, {Vale}, {Van
  Engelen}, {Van Lanen}, {Vavagiakis}, {Wagoner}, {Wang}, {Ward}, {Wollack},
  {Xu}, {Zago}, \& {Zhu}}]{choi2020}
{Choi}, S.~K., {Hasselfield}, M., {Ho}, S.-P.~P., {et~al.} 2020,
  \href{http://dx.doi.org/10.1088/1475-7516/2020/12/045}{\JournalTitle{\jcap},
  2020, 045}

\bibitem[{{Crisler} {et~al.}(2018){Crisler}, {Essig}, {Estrada}, {Fernandez},
  {Tiffenberg}, {Haro}, {Volansky}, {Yu}, \& {Sensei
  Collaboration}}]{sensei2018}
{Crisler}, M., {Essig}, R., {Estrada}, J., {et~al.} 2018,
  \href{http://dx.doi.org/10.1103/PhysRevLett.121.061803}{\JournalTitle{\prl},
  121, 061803}

\bibitem[{{Cyr-Racine} {et~al.}(2014){Cyr-Racine}, {de Putter}, {Raccanelli},
  \& {Sigurdson}}]{cyrracine2014}
{Cyr-Racine}, F.-Y., {de Putter}, R., {Raccanelli}, A., \& {Sigurdson}, K.
  2014,
  \href{http://dx.doi.org/10.1103/PhysRevD.89.063517}{\JournalTitle{\prd}, 89,
  063517}

\bibitem[{{Dvorkin} {et~al.}(2014){Dvorkin}, {Blum}, \&
  {Kamionkowski}}]{DvorkinBlum}
{Dvorkin}, C., {Blum}, K., \& {Kamionkowski}, M. 2014,
  \href{http://dx.doi.org/10.1103/PhysRevD.89.023519}{\JournalTitle{\prd}, 89,
  023519}

\bibitem[{{Fitzpatrick} \& {Zurek}(2010)}]{Fitzpatrick:2010br}
{Fitzpatrick}, A.~L., \& {Zurek}, K.~M. 2010,
  \href{http://dx.doi.org/10.1103/PhysRevD.82.075004}{\JournalTitle{\prd}, 82,
  075004}

\bibitem[{Gluscevic \& Boddy(2018)}]{GluscevicBoddy}
Gluscevic, V., \& Boddy, K.~K. 2018,
  \href{http://dx.doi.org/10.1103/PhysRevLett.121.081301}{\JournalTitle{\prl},
  121, 081301}

\bibitem[{{Green} {et~al.}(2019){Green}, {Amin}, {Meyers}, {Wallisch},
  {Abazajian}, {Abidi}, {Adshead}, {Ahmed}, {Ansarinejad}, {Armstrong},
  {Baccigalupi}, {Bandura}, {Barron}, {Battaglia}, {Baumann}, {Bechtol},
  {Bennett}, {Benson}, {Beutler}, {Bischoff}, {Bleem}, {Bond}, {Borrill},
  {Buckley-Geer}, {Burgess}, {Carlstrom}, {Castorina}, {Challinor}, {Chen},
  {Cooray}, {Coulton}, {Craig}, {Crawford}, {Cyr-Racine}, {D'Amico},
  {Demarteau}, {Dor{\'e}}, {Yutong}, {Dunkley}, {Dvorkin}, {Ellison}, {van
  Engelen}, {Escoffier}, {Essinger-Hileman}, {Fabbian}, {Filippini}, {Flauger},
  {Foreman}, {Fuller}, {Garcia}, {Garc{\'\i}a-Bellido}, {Gerbino}, {Gluscevic},
  {Gontcho}, {G{\'o}rski}, {Grin}, {Grohs}, {Gudmundsson}, {Hanany}, {Handley},
  {Hill}, {Hirata}, {Hlo{\v{z}}ek}, {Holder}, {Horiuchi}, {Huterer}, {Kadota},
  {Kamionkowski}, {Keeley}, {Khatri}, {Kisner}, {Kneib}, {Knox}, {Koushiappas},
  {Kovetz}, {L'Huillier}, {Lahav}, {Lattanzi}, {Lee}, {Liguori}, {Lin},
  {Loverde}, {Madhavacheril}, {Masui}, {McMahon}, {McQuinn}, {Meerburg},
  {Mirbabayi}, {Motloch}, {Mukherjee}, {Mun{\~o}z}, {Nagy}, {Newburgh},
  {Niemack}, {Nomerotski}, {Page}, {Piacentni}, {Pierpaoli}, {Pogosian},
  {Pryke}, {Puglisi}, {Stompor}, {Raveri}, {Reichardt}, {Rose}, {Rossi},
  {Ruhl}, {Schaan}, {Schubnell}, {Schutz}, {Sehgal}, {Senatore}, {Seo},
  {Sherwin}, {Simon}, {Slosar}, {Staggs}, {Stebbins}, {Suzuki}, {Switzer},
  {Timbie}, {Tristram}, {Trodden}, {Tsai}, {Umilt{\`a}}, {Di Valentino},
  {Vargas-Maga{\~n}a}, {Vieregg}, {Watson}, {Weiler}, {Whitehorn}, {Wu}, {Xu},
  {Xu}, {Yasini}, {Zaldarriaga}, {Zhao}, {Zhu}, \& {Zuntz}}]{green2019}
{Green}, D., {Amin}, M.~A., {Meyers}, J., {et~al.} 2019, \JournalTitle{\baas},
  51, 159

\bibitem[{{Hill} {et~al.}(2021){Hill}, {Calabrese}, {Aiola}, {Battaglia},
  {Bolliet}, {Choi}, {Devlin}, {Duivenvoorden}, {Dunkley}, {Ferraro},
  {Gallardo}, {Gluscevic}, {Hasselfield}, {Hilton}, {Hincks}, {Hlozek},
  {Koopman}, {Kosowsky}, {La Posta}, {Louis}, {Madhavacheril}, {McMahon},
  {Moodley}, {Naess}, {Natale}, {Nati}, {Newburgh}, {Niemack}, {Partridge},
  {Qu}, {Salatino}, {Schillaci}, {Sehgal}, {Sherwin}, {Sifon}, {Spergel},
  {Staggs}, {Storer}, {van Engelen}, {Vavagiakis}, {Wollack}, \&
  {Xu}}]{hill2021}
{Hill}, J.~C., {Calabrese}, E., {Aiola}, S., {et~al.} 2021, \JournalTitle{arXiv
  e-prints}, arXiv:2109.04451

\bibitem[{{Hlo{\v{z}}ek} {et~al.}(2017){Hlo{\v{z}}ek}, {Marsh}, {Grin},
  {Allison}, {Dunkley}, \& {Calabrese}}]{hlozek2017}
{Hlo{\v{z}}ek}, R., {Marsh}, D. J.~E., {Grin}, D., {et~al.} 2017,
  \href{http://dx.doi.org/10.1103/PhysRevD.95.123511}{\JournalTitle{\prd}, 95,
  123511}

\bibitem[{{Hooper} \& {Lucca}(2021)}]{2021arXiv211004024H}
{Hooper}, D.~C., \& {Lucca}, M. 2021, \JournalTitle{arXiv e-prints},
  arXiv:2110.04024

\bibitem[{{Lesgourgues}(2011)}]{class}
{Lesgourgues}, J. 2011, \JournalTitle{arXiv e-prints}, arXiv:1104.2932

\bibitem[{{Li} {et~al.}(2018){Li}, {Gluscevic}, {Boddy}, \&
  {Madhavacheril}}]{LiGluscevic2018}
{Li}, Z., {Gluscevic}, V., {Boddy}, K.~K., \& {Madhavacheril}, M.~S. 2018,
  \href{http://dx.doi.org/10.1103/PhysRevD.98.123524}{\JournalTitle{\prd}, 98,
  123524}

\bibitem[{{Maamari} {et~al.}(2021){Maamari}, {Gluscevic}, {Boddy}, {Nadler}, \&
  {Wechsler}}]{2021ApJ...907L..46M}
{Maamari}, K., {Gluscevic}, V., {Boddy}, K.~K., {Nadler}, E.~O., \& {Wechsler},
  R.~H. 2021,
  \href{http://dx.doi.org/10.3847/2041-8213/abd807}{\JournalTitle{\apjl}, 907,
  L46}

\bibitem[{{Marsh}(2016)}]{marsh2016}
{Marsh}, D. J.~E. 2016,
  \href{http://dx.doi.org/10.1016/j.physrep.2016.06.005}{\JournalTitle{\physrep},
  643, 1}

\bibitem[{{Nadler} {et~al.}(2020){Nadler}, {Wechsler}, {Bechtol}, {Mao},
  {Green}, {Drlica-Wagner}, {McNanna}, {Mau}, {Pace}, {Simon}, {Kravtsov},
  {Dodelson}, {Li}, {Riley}, {Wang}, {Abbott}, {Aguena}, {Allam}, {Annis},
  {Avila}, {Bernstein}, {Bertin}, {Brooks}, {Burke}, {Rosell}, {Kind},
  {Carretero}, {Costanzi}, {da Costa}, {De Vicente}, {Desai}, {Evrard},
  {Flaugher}, {Fosalba}, {Frieman}, {Garc{\'\i}a-Bellido}, {Gaztanaga},
  {Gerdes}, {Gruen}, {Gschwend}, {Gutierrez}, {Hartley}, {Hinton}, {Honscheid},
  {Krause}, {Kuehn}, {Kuropatkin}, {Lahav}, {Maia}, {Marshall}, {Menanteau},
  {Miquel}, {Palmese}, {Paz-Chinch{\'o}n}, {Plazas}, {Romer}, {Sanchez},
  {Santiago}, {Scarpine}, {Serrano}, {Smith}, {Soares-Santos}, {Suchyta},
  {Tarle}, {Thomas}, {Varga}, {Walker}, \& {DES Collaboration}}]{Nadler:2020}
{Nadler}, E.~O., {Wechsler}, R.~H., {Bechtol}, K., {et~al.} 2020,
  \href{http://dx.doi.org/10.3847/1538-4357/ab846a}{\JournalTitle{\apj}, 893,
  48}

\bibitem[{{Nadler} {et~al.}(2021){Nadler}, {Drlica-Wagner}, {Bechtol}, {Mau},
  {Wechsler}, {Gluscevic}, {Boddy}, {Pace}, {Li}, {McNanna}, {Riley},
  {Garc{\'\i}a-Bellido}, {Mao}, {Green}, {Burke}, {Peter}, {Jain}, {Abbott},
  {Aguena}, {Allam}, {Annis}, {Avila}, {Brooks}, {Carrasco Kind}, {Carretero},
  {Costanzi}, {da Costa}, {De Vicente}, {Desai}, {Diehl}, {Doel}, {Everett},
  {Evrard}, {Flaugher}, {Frieman}, {Gerdes}, {Gruen}, {Gruendl}, {Gschwend},
  {Gutierrez}, {Hinton}, {Honscheid}, {Huterer}, {James}, {Krause}, {Kuehn},
  {Kuropatkin}, {Lahav}, {Maia}, {Marshall}, {Menanteau}, {Miquel}, {Palmese},
  {Paz-Chinch{\'o}n}, {Plazas}, {Romer}, {Sanchez}, {Scarpine}, {Serrano},
  {Sevilla-Noarbe}, {Smith}, {Soares-Santos}, {Suchyta}, {Swanson}, {Tarle},
  {Tucker}, {Walker}, {Wester}, \& {DES Collaboration}}]{2021PhRvL.126i1101N}
{Nadler}, E.~O., {Drlica-Wagner}, A., {Bechtol}, K., {et~al.} 2021,
  \href{http://dx.doi.org/10.1103/PhysRevLett.126.091101}{\JournalTitle{\prl},
  126, 091101}

\bibitem[{{Nguyen} {et~al.}(2021){Nguyen}, {Sarnaaik}, {Boddy}, {Nadler}, \&
  {Gluscevic}}]{2021PhRvD.104j3521N}
{Nguyen}, D.~V., {Sarnaaik}, D., {Boddy}, K.~K., {Nadler}, E.~O., \&
  {Gluscevic}, V. 2021,
  \href{http://dx.doi.org/10.1103/PhysRevD.104.103521}{\JournalTitle{\prd},
  104, 103521}

\bibitem[{{Prince} \& {Dunkley}(2019)}]{princedunkley}
{Prince}, H., \& {Dunkley}, J. 2019,
  \href{http://dx.doi.org/10.1103/PhysRevD.100.083502}{\JournalTitle{\prd},
  100, 083502}

\bibitem[{{Revels} {et~al.}(2016){Revels}, {Lubin}, \&
  {Papamarkou}}]{forwarddiff}
{Revels}, J., {Lubin}, M., \& {Papamarkou}, T. 2016, \JournalTitle{arXiv
  e-prints}, arXiv:1607.07892

\bibitem[{{Rogers} {et~al.}(2021){Rogers}, {Dvorkin}, \&
  {Peiris}}]{2021arXiv211110386R}
{Rogers}, K.~K., {Dvorkin}, C., \& {Peiris}, H.~V. 2021, \JournalTitle{arXiv
  e-prints}, arXiv:2111.10386

\bibitem[{{Simons Observatory Collaboration}(2019)}]{so_forecast:2019}
{Simons Observatory Collaboration}. 2019,
  \href{http://dx.doi.org/10.1088/1475-7516/2019/02/056}{\JournalTitle{Journal
  of Cosmology and Astro-Particle Physics}, 2019, 056}

\bibitem[{{Slatyer} \& {Wu}(2018)}]{SlatyerWu}
{Slatyer}, T.~R., \& {Wu}, C.-L. 2018,
  \href{http://dx.doi.org/10.1103/PhysRevD.98.023013}{\JournalTitle{\prd}, 98,
  023013}

\bibitem[{{Smith} {et~al.}(2003){Smith}, {Peacock}, {Jenkins}, {White},
  {Frenk}, {Pearce}, {Thomas}, {Efstathiou}, \& {Couchman}}]{halofit}
{Smith}, R.~E., {Peacock}, J.~A., {Jenkins}, A., {et~al.} 2003,
  \href{http://dx.doi.org/10.1046/j.1365-8711.2003.06503.x}{\JournalTitle{\mnras},
  341, 1311}

\bibitem[{{Thornton} {et~al.}(2016){Thornton}, {Ade}, {Aiola}, {Angil{\`e}},
  {Amiri}, {Beall}, {Becker}, {Cho}, {Choi}, {Corlies}, {Coughlin}, {Datta},
  {Devlin}, {Dicker}, {D{\"u}nner}, {Fowler}, {Fox}, {Gallardo}, {Gao},
  {Grace}, {Halpern}, {Hasselfield}, {Henderson}, {Hilton}, {Hincks}, {Ho},
  {Hubmayr}, {Irwin}, {Klein}, {Koopman}, {Li}, {Louis}, {Lungu}, {Maurin},
  {McMahon}, {Munson}, {Naess}, {Nati}, {Newburgh}, {Nibarger}, {Niemack},
  {Niraula}, {Nolta}, {Page}, {Pappas}, {Schillaci}, {Schmitt}, {Sehgal},
  {Sievers}, {Simon}, {Staggs}, {Tucker}, {Uehara}, {van Lanen}, {Ward}, \&
  {Wollack}}]{thornton/2016}
{Thornton}, R.~J., {Ade}, P.~A.~R., {Aiola}, S., {et~al.} 2016,
  \href{http://dx.doi.org/10.3847/1538-4365/227/2/21}{\JournalTitle{\apjs},
  227, 21}

\bibitem[{{Torrado} \& {Lewis}(2021)}]{cobaya}
{Torrado}, J., \& {Lewis}, A. 2021,
  \href{http://dx.doi.org/10.1088/1475-7516/2021/05/057}{\JournalTitle{\jcap},
  2021, 057}

\bibitem[{Xu {et~al.}(2018)Xu, Dvorkin, \& Chael}]{Xu189710}
Xu, W.~L., Dvorkin, C., \& Chael, A. 2018,
  \href{http://dx.doi.org/10.1103/PhysRevD.97.103530}{\JournalTitle{\prd}, 97,
  103530}

\end{thebibliography}


\appendix
\FloatBarrier
\label{appendix}

\begin{table}[ht]
\centering
\centering
\begin{tabular}{ccccc}
\hline\hline
$n$ & $m_\chi$  & \planck{} & ACT + \planck{} & A+P electron \\ 
    &  [GeV] & [95\%, cm$^2$] & [95\%, cm$^2$] & [95\%, cm$^2$] \\ \hline
$-4$ & 0.001 & $1.3 \times 10^{-41}$ & $1.8 \times 10^{-41}$ & $1.4 \times 10^{-37}$ \\
$-4$ & 0.01 & $1.4 \times 10^{-41}$ & $1.9 \times 10^{-41}$ & $1.4 \times 10^{-36}$ \\
$-4$ & 0.1 & $1.5 \times 10^{-41}$ & $2.0 \times 10^{-41}$ & $9.1 \times 10^{-36}$ \\
$-4$ & 1.0 & $2.7 \times 10^{-41}$ & $3.7 \times 10^{-41}$ & $9.1 \times 10^{-35}$ \\
$-4$ & 10.0 & $1.6 \times 10^{-40}$ & $2.1 \times 10^{-40}$ & $8.0 \times 10^{-34}$ \\
$-4$ & 100.0 & $1.4 \times 10^{-39}$ & $1.9 \times 10^{-39}$ & $8.4 \times 10^{-33}$ \\
$-4$ & 1000.0 & $1.4 \times 10^{-38}$ & $1.9 \times 10^{-38}$ & $5.3 \times 10^{-32}$ \\ \hline
$-2$ & 0.001 & $1.8 \times 10^{-33}$ & $2.3 \times 10^{-33}$ & $7.0 \times 10^{-32}$ \\
$-2$ & 0.01 & $1.8 \times 10^{-33}$ & $2.2 \times 10^{-33}$ & $5.2 \times 10^{-31}$ \\
$-2$ & 0.1 & $2.0 \times 10^{-33}$ & $2.4 \times 10^{-33}$ & $4.9 \times 10^{-30}$ \\
$-2$ & 1.0 & $3.6 \times 10^{-33}$ & $4.6 \times 10^{-33}$ & $5.7 \times 10^{-29}$ \\
$-2$ & 10.0 & $2.1 \times 10^{-32}$ & $2.5 \times 10^{-32}$ & $5.6 \times 10^{-28}$ \\
$-2$ & 100.0 & $1.9 \times 10^{-31}$ & $2.4 \times 10^{-31}$ & $4.9 \times 10^{-27}$ \\
$-2$ & 1000.0 & $1.9 \times 10^{-30}$ & $2.4 \times 10^{-30}$ & $5.8 \times 10^{-26}$ \\ \hline
0 & 0.001 & $5.7 \times 10^{-26}$ & $3.5 \times 10^{-26}$ & $6.5 \times 10^{-27}$ \\
0 & 0.01 & $1.1 \times 10^{-25}$ & $6.4 \times 10^{-26}$ & $4.0 \times 10^{-26}$ \\
0 & 0.1 & $1.9 \times 10^{-25}$ & $1.1 \times 10^{-25}$ & $6.4 \times 10^{-25}$ \\
0 & 1.0 & $4.7 \times 10^{-25}$ & $2.9 \times 10^{-25}$ & $4.2 \times 10^{-24}$ \\
0 & 10.0 & $2.9 \times 10^{-24}$ & $1.9 \times 10^{-24}$ & $5.7 \times 10^{-23}$ \\
0 & 100.0 & $2.9 \times 10^{-23}$ & $1.8 \times 10^{-23}$ & $1.1 \times 10^{-21}$ \\
0 & 1000.0 & $2.9 \times 10^{-22}$ & $1.8 \times 10^{-22}$ & $3.7 \times 10^{-21}$ \\ \hline
\end{tabular} \quad
\begin{tabular}{ccccc}
\hline\hline
$n$ & $m_\chi$  & \planck{} & ACT + \planck{} & A+P electron \\ 
    &  [GeV] & [95\%, cm$^2$] & [95\%, cm$^2$] & [95\%, cm$^2$] \\ \hline
2 & 0.001 & $3.9 \times 10^{-21}$ & $5.3 \times 10^{-21}$ & $1.5 \times 10^{-22}$ \\
2 & 0.01 & $5.2 \times 10^{-20}$ & $7.0 \times 10^{-20}$ & $9.3 \times 10^{-22}$ \\
2 & 0.1 & $6.5 \times 10^{-19}$ & $8.5 \times 10^{-19}$ & $1.1 \times 10^{-20}$ \\
2 & 1.0 & $8.7 \times 10^{-18}$ & $7.3 \times 10^{-18}$ & $1.1 \times 10^{-19}$ \\
2 & 10.0 & $9.2 \times 10^{-17}$ & $6.3 \times 10^{-17}$ & $1.4 \times 10^{-18}$ \\
2 & 100.0 & $9.5 \times 10^{-16}$ & $6.0 \times 10^{-16}$ & $1.2 \times 10^{-17}$ \\
2 & 1000.0 & $9.7 \times 10^{-15}$ & $5.8 \times 10^{-15}$ & $9.2 \times 10^{-17}$ \\ \hline
4 & 0.001 & $1.1 \times 10^{-16}$ & $1.9 \times 10^{-16}$ & $1.4 \times 10^{-18}$ \\
4 & 0.01 & $1.4 \times 10^{-14}$ & $2.2 \times 10^{-14}$ & $1.8 \times 10^{-17}$ \\
4 & 0.1 & $1.4 \times 10^{-12}$ & $2.3 \times 10^{-12}$ & $1.9 \times 10^{-16}$ \\
4 & 1.0 & $7.3 \times 10^{-11}$ & $1.1 \times 10^{-10}$ & $1.7 \times 10^{-15}$ \\
4 & 10.0 & $1.3 \times 10^{-9}$ & $1.5 \times 10^{-9}$ & $1.7 \times 10^{-14}$ \\
4 & 100.0 & $1.3 \times 10^{-8}$ & $1.6 \times 10^{-8}$ & $1.5 \times 10^{-13}$ \\
4 & 1000.0 & $8.9 \times 10^{-10}$ & $3.5 \times 10^{-10}$ & $1.3 \times 10^{-12}$ \\ \hline
6 & 0.001 & $2.3 \times 10^{-12}$ & $3.8 \times 10^{-12}$ & $7.4 \times 10^{-15}$ \\
6 & 0.01 & $2.5 \times 10^{-9}$ & $4.3 \times 10^{-9}$ & $7.3 \times 10^{-14}$ \\
6 & 0.1 & $2.2 \times 10^{-6}$ & $4.0 \times 10^{-6}$ & $1.2 \times 10^{-12}$ \\
6 & 1.0 & $5.7 \times 10^{-4}$ & $9.7 \times 10^{-4}$ & $1.8 \times 10^{-11}$ \\
6 & 10.0 & $1.3 \times 10^{-2}$ & $1.6 \times 10^{-2}$ & $1.2 \times 10^{-10}$ \\
6 & 100.0 & $8.2 \times 10^{-3}$ & $8.4 \times 10^{-3}$ & $1.1 \times 10^{-9}$ \\
6 & 1000.0 & $1.6 \times 10^{-3}$ & $3.8 \times 10^{-4}$ & $1.1 \times 10^{-8}$ \\ \hline
\end{tabular}
  \caption{Upper 95\% limits on DM cross sections, sampled with linear prior, for models with velocity dependence $n$ and DM mass $m_{\chi}$, for proton scattering (column 3 \& 4) and electron scattering (column 5). Many of the models with $n \neq 0$ do not show an improvement in the upper bound over Planck with the addition of ACT DR4, due to the nonzero best-fitting cross sections discussed in Section \ref{subsec:veldepconstr}.} 
  \label{table:constraints}
\end{table}

In this appendix we estimate how much statistical power ACT contributes to the ACT+\emph{Planck} combination, relative to \emph{Planck} alone. The constraining power is distinct from the upper bound, since the upper bound is sensitive to statistical fluctuations which change the peaks of the respective likelihoods. We perform a simple Fisher matrix analysis with the likelihoods directly.\footnote{ \url{https://xzackli.github.io/realfisher/fisheranalysis.jl.html}.} The Fisher matrix for parameters $\{\theta_i\}$ is the expectation value for the Hessian of the log-likelihood $\mathcal{L}$,
\begin{equation}
F_{ij} = - \left\langle\frac{ \partial^2 \mathcal{L}}{\partial \theta_i \, \partial \theta_j} \right\rangle_{\theta}.
\end{equation}

We approximate this expectation value with the value at the peak of the likelihood. To compute the Hessian of the log-likelihoods, we use forward-mode automatic differentiation (AD) in the \texttt{ForwardDiff.jl} package \citep{forwarddiff} within the Julia language \citep{bezanson2017julia}. Although it is possible to derive an analytic Hessian, the ACT likelihood has complexities (binning, nontrivial bandpower window functions, deep and wide patches) which would make this tedious.  We re-implement the ACT and \emph{Planck} likelihood in Julia, to enable the use of this AD package. Our new implementation reproduces the ACT DR4 likelihood \citep{aiola2020} to numerical precision. We use the compressed high-$\ell$ likelihood provided in \cite{princedunkley} for \emph{Planck} 2018, and we reproduce this likelihood to numerical precision as well. We obtain gradients of the model spectra using finite differences.

After computing the negative Hessian of the log-likelihood numerically with AD, we invert it to obtain the covariance matrix via the Cramer-Rao bound, $C_{ij} = F_{ij}^{-1}$. 
To represent the combined likelihoods, we add the Fisher matrices corresponding to ACT alone and \emph{Planck} alone. We also impose a prior on the error of the optical depth to reionization, $\sigma(\tau_{\mathrm{reio}}) = 0.015$, to replace the large-scale polarization data in \emph{Planck} by adding $(0.015)^2$ to the diagonal element of the Fisher matrix corresponding to $\tau_{\mathrm{reio}}$. 

This Fisher analysis configuration reproduces the $\Lambda$CDM constraints in \cite{aiola2020} fairly well, with only 10$-$20\% differences for each of the parameter errors in each configuration (\emph{Planck}, ACT, ACT and \emph{Planck} combined).

For parameter $\theta_i$, we then compute the marginalized error $\sigma(\theta_i) = \sqrt{C_{ii}}$.
We also define an overall figure of merit (FoM) that describes the full extension model, $\mathrm{FoM} = 1 / \sqrt{\det F}$.
We present ratios of these quantities in Table \ref{table:fisher} for the \emph{Planck} likelihood alone (P), with respect to ACT and \emph{Planck} combined (AP). We use the ACT and \emph{Planck} combined best-fit as the fiducial model. We perform this analysis for the 1\,GeV case, as the degeneracy between mass and cross section results in similar results for other masses. Although the overall figure of merit improves by roughly a factor of two across all models, the ACT data improves the constraint on $\sigma_{0}$ by 30$-$40\% for $n\geq0$ models. For $n<0$ models, we see that the ACT data contributes little constraining power. 

\begin{table}[t]
\centering
\caption{Statistical constraining power estimates from \planck{} alone (P), and ACT and \planck{} combined (AP), for $m_{\chi} = 1 \text{ GeV}$.}
\label{table:fisher}
\begin{tabular}{ccc}
\hline\hline
$n$ & $\sigma^{\rm P}(\sigma_{0}) / \sigma^{\rm AP}(\sigma_{0})$ & ${\rm FoM}^{\rm P} / {\rm FoM}^{\rm AP}$ \\ \hline
-4 & 1.07 & 1.64 \\ 
-2 & 1.06 & 1.62 \\ 
0 & 1.34 & 2.06 \\ 
2 & 1.49 & 2.29 \\ 
4 & 1.32 & 2.04 \\ 
6 & 1.40 & 2.15 \\ 
\hline
\end{tabular}
\end{table}

\end{document}